\documentclass{aa}  

\usepackage{graphicx}
\usepackage{longtable}

\usepackage{lscape}
\usepackage{eqnarray,amsmath}
\usepackage{txfonts}
\usepackage{xcolor}

\usepackage{rotating}
\usepackage{pdflscape}
\usepackage{tabulary} 
\usepackage{hyperref}
\usepackage{tablefootnote}
\usepackage{subfig}

\begin{document} 

\title{Modified models of radiation pressure instability in application to 10, 10$^5$,  and 10$^7$ $M_{\odot}$ accreting black holes}
\titlerunning{Timescales in radiation pressure dominated disks}

   \author{Marzena Śniegowska
          \inst{1,2}\fnmsep\thanks{msniegowska@camk.edu.pl}
          \and
          Mikołaj Grzędzielski\inst{2}
           \and
          Bo\.zena Czerny\inst{2}
           \and
          Agnieszka Janiuk\inst{2}
          }

              \institute{Nicolaus Copernicus Astronomical Center (PAN), ul. Bartycka 18, 00-716 Warsaw, Poland\\
         \and Center for Theoretical Physics, Polish Academy of Sciences, Al. Lotnik\'ow 32/46, 02-668 Warsaw, Poland\\}


 
  \abstract
   {Some of the accreting black holes exhibit much stronger variability patterns than the usual stochastic variations. Radiation pressure instability is one of the proposed mechanisms which could account for this effect. }
   {We aim to model luminosity changes for objects with black hole mass of 10, 10$^5$,  and 10$^7$ solar masses, using the time-dependent evolution of an accretion disk unstable due to the dominant radiation pressure. We explore the influence of the hot coronal flow above the cold disk, the inner purely hot flow,  and the effect of magnetic field on the time evolution of disk-corona system. In the case of Intermediate Mass Black Holes and AGN we also explore the role of the disk outer radius, motivated by the fact that the disk fed by Tidal Disruption Events can be quite small in size.}
   {We use a 1-dimensional, vertically integrated time-dependent numerical scheme which models simultaneous evolution of the disk and corona, coupled by the vertical mass exchange. We parameterize the strength of large-scale toroidal magnetic fields according to a local accretion rate. We also discuss the possibility of presence of an inner optically thin flow, namely the Advection-Dominated Accretion Flow (ADAF) which requires modification of the inner boundary condition of the cold disk flow. For the set of the global parameters we calculate the variability timescales and outburst amplitudes of the disk and the corona. 
}
   {We found that the role of the inner ADAF and the accreting corona are relatively unimportant but the outburst character strongly depends on the magnetic field and the outer radius of the disk if this radius is smaller (due to TDE phenomenon) than the size of the instability zone in a stationary disk with infinite radius. For microquasars, the dependence on the magnetic field is monotonic, and the period decreases with the field strength. For larger black hole masses, the dependence is non-monotonic, and initial rise of the period is later replaced with the relatively rapid decrease as the magnetic field continues to rise. Still stronger magnetic field stabilizes the disk. Assumption of the smaller disk outer radius considered for 10$^5$,  and 10$^7 M_{\odot}$ shortens the outbursts and for some parameter range leads to complex multi-scale outbursts thus approaching the deterministic chaos behaviour. }
   {Our computations confirm that the radiation pressure instability model can account for heartbeat states in microquasars. Rapid variability detected in IMBH in the form of Quasi-Periodic Ejection can be consistent with the model but only if combined with TDE phenomenon. Yearly repeating variability in Changing Look AGN also requires, in our model, small outer radius either due to the recent TDE or due to the presence of the gap in the disk related to the presence of a secondary black hole. }


   \maketitle
%

\section{Introduction}
The variability around black holes across their mass range shows different patterns of the amplitudes, shapes of the lightcurves, and timescales. 
The interpretation of the physical nature of those changes is difficult. Part of the variability is stochastic in nature, and these stochastic variations are seen both in X-rays and in the optical band  \citep[e.g.][]{lehto1993,czerny1999,gaskell2003,2021MNRAS.508.3975K}. However, in some sources, the variations are much stronger than the usual stochastic changes. We do not consider here well-known 
state transitions
in X-ray binary systems, which happen in timescales of days, corresponding to the viscous evolution of the outer disk, but much more rapid, sometimes regular changes of the source luminosity. 

For example, in the case of microquasars (which black hole masses are in the range of 5 - 20 $M_{\odot}$) changes appear on timescales of tens to hundreds of seconds. Regular quasi-periodic outbursts with the period depending on the mean flux were observed in GRS 1915+105 \citep{belloni2000}. This regular variability was well modeled by the radiation pressure instability in the central parts of the accretion disk \citep{pringle73,le74,2000nayakshin,2000janiuk,2002janiuk}. Similar 'heartbeat' variability of IGR J17091 microquasar (RA: 01h19m08.68s  DEC: -34d11m30.5s after NED\footnote{https://ned.ipac.caltech.edu/classic/}) was discovered by \cite{2011Altamirano}, and its luminosity changes in X-ray band have also been modeled by \cite{2015janiuk} by the accretion disk radiation pressure instability. However, recently this source returned to the quiescent state \citep{2020Pereyra}. On the other hand, \cite{2015Bagnoli} discovered the \textit{heartbeat} states in MXB 1730-335 (the `Rapid Burster'), later explored by \citet{Maselli2018}.

Interesting semi-periodic variations were also discovered in the source HLX-1 located in the galaxy ESO 243-49. This source most likely contains an intermediate mass black hole of $10^5 M_{\odot}$ and it underwent recurrent outbursts on the timescale of 400 days \citep{2015Yan}. The behavior of this source was also well modeled by the radiation pressure instability \citep{2016Wu}.

For more massive sources like active galactic nuclei (in which the typical mass of the central black hole is around 10$^7$ M$_{\odot}$), the phenomenon which goes well beyond the stochastic variability is the occasional change of their spectral state. The typical observed timescales of such changes are rather counted in years \citep[see introductions to recent papers, like][]{yang2018,noda2018,trakhtenbrot2019,graham2020,oknyansky2019,2020A&A...641A.167S}.
Changes in Changing Look (CL) AGN may be caused by more than one mechanism which opens the possibility to model and explore different scenarios for these phenomena.
Several mechanisms are presented in the literature. \cite{ross2018} for the J1100-0053 source proposed the instability in a cold disk/ADAF caused by magnetic fields threading the inner disk. \citet{noda2018} discussed the temporary disappearance of the warm corona for Mrk 1018. 
For the same object, \cite{2021ApJ...916...61F} proposed magnetic accretion disk outflows, and modelled UV/optical spectral shape of Mrk 1018, consistently reproducing the observed changes.
Supermassive black hole binaries, which may cause tidal interaction between disks leading to state-changes, are suggested by \citet{2020wangbon} as one of the possible scenarios. \cite{2021scepi} proposed that in the source 1ES 1927+654 the CL event results from reversing magnetic field. This is motivated by analogy with the 11-year solar cycle, and the field reversals and changes in accretion rate were indeed seen in MHD simulations of magnetically-arrested disks.  \citet{2021raj} performed simulation of the tearing disk structure, which may cause instabilities of the inner accretion flow \citep[see also][and references therein]{2021ApJ...909...81R}.

In the current paper, we concentrate on the cases when the CL phenomenon repeats in a given source. Such quasi-periodic behavior rules out TDE or obscuration events. 
NGC 1566 is the example of CL AGN with semi-regular outbursts observed in the optical band \citep{alloin1986, oknyansky2019}. If we assume thermal/viscous instability is led by radiation pressure, we can mimic the outburst cycle.
In \cite{2020A&A...641A.167S} we proposed the simple time-dependent toy model of accretion disk under radiation pressure instability, in which the timescale of outbursts is regulated by the thickness of the unstable zone. \cite{2021ApJ...910...97P} presented the extension of \cite{2020A&A...641A.167S} model by adding a magnetic field and showed that the timescale of outbursts can be significantly reduced by the magnetic fields. 
The possibility of shortening timescales in CL AGN phenomena by magnetization of accretion disk was also suggested by \citet{dexter2019}.

Recently, a new, intriguing class of variable objects has been found. The first one was
GSN 069  \citep{Miniutti2019}.
This source is characterized by short and symmetric repetitive flares (every nine hours). Four more sources of similarly short eruption timescales have been found, and they are now known as Quasi-Periodic Ejection events (RX J1301.9+2747, \citealt{giustini2020}; eRO-QPE1 and eRO-QPE, \citealt{arcodia2021}; 2MASXJ0249, \citealt{2021Chakraborty}). It is not clear whether they are related to CL AGN, 
\cite{2020king} suggests that this is a TDE event nearly missed. Similarly,
\citep{2021xian} suggests that those outbursts may be driven by star-disk collisions, and \citep{2021arXiv210903471Z} shows with stellar evolution code MESA that hydrogen-deficient stars are good candidates for this scenario. \citet{sukova2021} proposed periodic plasmoid ejection by in-spiralling star. 
\cite{2021MNRAS.503.1703I} proposes the scenario with a self-lensing of the binary SMBH, while
\cite{2021Metzger} propose the QPE mechanism based on orbital co-planarity of stars where at least one of them overflows its Roche lobe and accretes onto the SMBH.
For those 4 sources the lack of broad emission lines in the optical band is also characteristic, and this feature rules out the possibility to estimate black hole masses independently using broad line diagnostics. As \citet{2022A&A...659L...2W} point out, emission lines noticed in those sources are typical of star-forming or accreting galaxies.

Our goal in this paper is to obtain a grid of models based on radiation pressure instability, with some modifications, and to see which of the observed timescales in different types of objects are consistent with the radiation pressure instability mechanism as the driving factor.

To perform this work we use the time-dependent code GLADIS (Global Accretion Disk Instability Simulation) developed originally by \cite{2002janiuk} and now publicly available \citep{2020mbhe.confE..48J}. We model the accretion disk evolution by solving the equatorial disk temperature and the surface density time dependent equations, under the assumption of vertical hydrostatic equilibrium.
We performed preliminary simulations for 10$^7$M$_{\odot}$ using this model in \cite{sniegowska2022}. However, the obtained timescale of outbursts was too long (more than 100 years) in comparison to observed CL AGN. Thus, we decided to explore possible additional mechanisms which may shorten the period to a few years' timescale. 

In this work, we extend the model in GLADIS by adding a new boundary condition in order to represent correctly the inner ADAF, and this is done by assuming a constant inflow of mass from the corona to the inner ADAF at the transition radius. 
We also account for an extra cooling component based on the idea of a 'dead zone' \citep{2007MNRAS.375.1070B, 2015begelman} to shrink the time scale of outbursts without the damping effect. Finally, we study the potential role of TDE by allowing the disk outer radius to be much smaller than the full unstable zone in infinite accretion disk \citep[for these ranges, see][]{janiuk2011}.

We explore the properties of the model for $10 M_\odot$, $10^5 M_\odot$ and $10^7 M_\odot$ black hole mass cases, which correspond to a microquasar, intermediate-mass source GSN 069 with QPEs, and the galaxy NGC 1566 with the cyclic outbursts.




\section{Model}

We calculate the time evolution of the disk in a two-zone approximation of the vertical structure of the disk/coronal flow using the code GLADIS \citep{2020mbhe.confE..48J}. The disk structure and the corona structure at each radius are separately vertically averaged, as in the classical description of the disk \citep{ss73} or ADAF flow \citep{ichimaru77,1994ApJ...428L..13N} but there is a mass exchange between the two, parameterized as in Equation (10), CASE (b) of \citet{2007janiuk}. We thus include the disk evaporation to the corona, in a way dependent on the disk thermal state.
We consider two options for the viscous dissipation and angular momentum transfer: viscous torque proportional to the total pressure, $\alpha P_{tot}$, and viscous torque parameterized by the geometrical mean between the total pressure and the gas pressure, $\alpha \sqrt{P_{tot}P_{gas}}$ \citep{janiuk2011}. We, therefore, do not discuss here even more general parametrization by the $\mu$ coefficient, as in \citet{szuszkiewicz1990} and \citet{grzedzielski2017}. 

\begin{figure}
    \centering
    \includegraphics[scale=0.35]{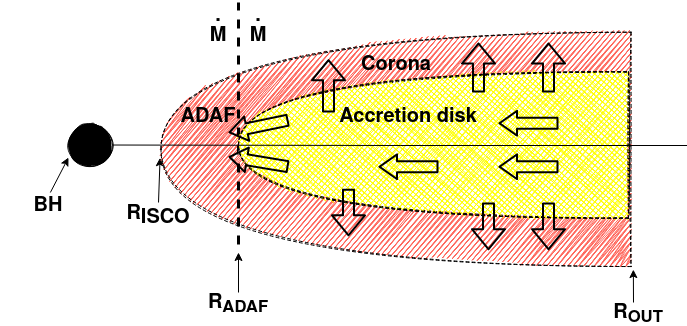}
    \caption{Schematic view of the innermost part of the flow: accretion disk (yellow), hot corona above the disk (red), which transforms into ADAF. The horizontal black line represents the equatorial plane. The vertical dashed line represents the boundary condition between ADAF and outer flow.  We perform computations between R$_{ADAF}$ and R$_{out}$}. 
    \label{fig:our_model}
\end{figure}

However, in comparison with previous code applications \citep{2000janiuk,2002janiuk,janiuk2011,2015janiuk,grzedzielski2017}, we allow for three major modifications:
\begin{itemize}
\item the development of the inner ADAF zone 
\item the modification of the disk vertical structure by the magnetic field
\item the decrease of the disk outer radius.
\end{itemize}
The formation of the inner ADAF means that the cold disk disappears and the proper boundary conditions must be formulated to match the ADAF and the coronal flow in the outer disk. We illustrate schematically the geometry of the model in Figure~\ref{fig:our_model}. Setting the outer disk radius at the arbitrary value in the case of AGN is motivated by the fact that CL behaviour might be related to TDE, and in this case, assuming a large enough outer radius of the disk to cover the whole instability zone, as was done it the papers cited above, may not be justified.


\subsection{Boundary condition between ADAF and outer flow}
\label{sect:ADAF}
We assume that the transition radius between the inner ADAF and the outer two-zone disk/corona flow is set by the disappearance of the disk. The flow in the inner ADAF is stable, no mass accumulation is expected there. Therefore, since the disk does not transfer mass anymore at the transition radius, the continuity equation should be satisfied between the inner ADAF and the corona. The proper inner condition implies the constraints for the three innermost points of the radial grid since the equations are of the second order. Point 0 is the innermost grid point where only ADAF is present which transfers all the material from the corona. We assume
   \begin{equation}
   \dot M_{0}^{cor} = \dot M_{1}^{cor}
      \label{eq:zalozenie}
   \end{equation}
for the coronal flow. Since  $\dot M^{cor} = -2\pi r\Sigma^{cor} V_r^{cor}$ (see Equation 27 in \citealt{2002janiuk}), where $\Sigma^{cor}$ is the corona surface density and $V_r^{cor}$ is the radial velocity, applying this equation to the condition above we get
\begin{equation}
\Sigma_0^{cor}  r_0 V_{r0}^{cor}= \Sigma_1^{cor} r_1 V_{r1}^{cor}.
 \label{eq:zalozenie_rozpisane}
\end{equation}

The radial velocity in the corona is calculated as in \citet{2007janiuk}:
 \begin{equation}
   V_{r}^{cor} = \frac{3}{\Sigma_{cor}r^{1/2}}\frac{\partial}{\partial r}(\nu_{cor}\Sigma_{cor}r^{1/2}),
   \end{equation}
where $\nu_{cor}$ is the kinematic viscosity of the corona.

If we consider three points $r_0, r_1, r_2$ and approximate derivatives as  differences between two points $\Delta r_{01}$ for the section $r_1 - r_0 $ and $\Delta r_{12}$  for the section $r_2 - r_1 $ in the grid we get the following expression for velocity:
 \begin{equation}
   V_{r0}^{cor} = \frac{3}{\Sigma_{0^{cor}}r_0^{1/2}} \left( \frac{\nu_1 \Sigma_1^{cor} r_1^{1/2} - \nu_0 \Sigma_0^{cor} r_0^{1/2}}{\Delta r_{01}}\right)
   \end{equation}
 for the left part of eq. \ref{eq:zalozenie_rozpisane} and:
    \begin{equation}
   V_{r1}^{cor} = \frac{3}{\Sigma_{1}^{cor}r_1^{1/2}} \left( \frac{\nu_2 \Sigma_2^{cor} r_2^{1/2} - \nu_1 \Sigma_1^{cor} r_1^{1/2}}{\Delta r_{12}}\right)
   \end{equation}
   for the right one.
Using velocity expressions we get:
 \begin{equation}
  \resizebox{.95\hsize}{!}{
    $\Sigma_0^{cor}  r_0 \frac{3}{\Sigma_{0}^{cor}r_0^{1/2}} \left( \frac{\nu_1 \Sigma_1^{cor} r_1^{1/2} - \nu_0 \Sigma_0^{cor} r_0^{1/2}}{\Delta r_{01}}\right) = \Sigma_1^{cor} r_1 \frac{3}{\Sigma_{1}^{cor}r_1^{1/2}} \left( \frac{\nu_2 \Sigma_2^{cor} r_2^{1/2} - \nu_1 \Sigma_1^{cor} r_1^{1/2}}{\Delta r_{12}}\right)$ }
   \end{equation}
and after simplification, we find the requested inner boundary condition for $\Sigma_0$ as:
  \begin{equation}
    \Sigma_0^{cor} = \frac{r_1^{1/2}}{\nu_0}  \left[\frac{\nu_1\Sigma_1^{cor}}{r_0^{1/2}} + \frac{\Delta r_{01}}{r_0 \Delta r_{12}} \left( r_1^{1/2}\nu_1\Sigma_1^{cor} - r_2^{1/2}\nu_2\Sigma_2^{cor}\right) \right].
   \end{equation}
The values of $\Sigma_1^{cor}$ and $\Sigma_2^{cor}$ are calculated from the time evolution equations. 
The inner ADAF thus transfers all the material arriving through the corona to the innermost zone. This value of the surface density is time-dependent, derived from the quantities known in each new time step from time-dependent evolutionary equations. This boundary condition is used only when the inner radius is different than 3R$_{schw}$.
\subsection{The role of the magnetic field and the appearance of the 'dead zone'}
\label{sec:modela_modelb}
In the standard accretion disk theory \citet{ss73} the magnetic field appears as a provider of the viscosity mechanism, parameterized through $\alpha$. The actual mechanism behind is the magnetorotational instability (MRI) \citep{balbus1991}. Its action has been seen in numerous MHD simulations \citep[e.g.][]{fromang2007,Jiang2014,pjanka2020} as well as in experimental studies in the laboratory \citep[see e.g.][and the references therein]{winarto2020}. However, as discussed by \citet{2015begelman}, MRI can lead to the generation of a toroidal magnetic field of considerable strength, and a dead zone can appear in the middle of the disk vertical structure, where MRI ceases to operate but magnetic energy continues to flow upward. This modifies considerably the effective vertical structure.

The criterium for 'dead zone' development from \cite{2007MNRAS.375.1070B} (Equation 7 therein; see also the discussion in \citealt{2015begelman})
\begin{equation}
v_{A}^2 \gtrsim c_s v_K, \\
\end{equation}
where  v$_{A}$ represents the  Alfv\'en speed, c$_s$ represents the speed of sound, and 
$v_A^2 = P_{tot}/\rho$ and $c_s^2 = P_{gas}/\rho$. Thus the criterion for 'dead zone' formation is
\begin{equation}
 \frac{P_{tot}}{\rho} \gtrsim v_K \sqrt{\frac{P_{gas}}{\rho}}.\\
\label{eq:criterium}
\end{equation}
We can evaluate when this criterion is satisfied using the 
hydrostatic equilibrium
\begin{equation}
   \frac{P_{tot}}{\rho} = v_K^2 \left( \frac{H}{r}\right)^2.
\end{equation}
We have two most extreme possible cases, gas pressure dominance, and the radiation pressure dominance.\\

i) Dominance of P$_{gas}$, P$_{tot}$ = P$_{gas}$\\

In this case we can rewrite the Equation \ref{eq:criterium} as
\begin{equation}
 \frac{P_{gas}}{\rho} \gtrsim v_K \sqrt{\frac{P_{gas}}{\rho}}\\
\label{eq:critpgas}
\end{equation}
and combining it with the hydrostatic equilibrium the criterion for the dead zone development reduces to
\begin{equation}
\left( \frac{H}{r}\right) \gtrsim 1.
 \end{equation}
 which is never satisfied. Thus, for a gas-dominated solution the magnetic field does not modify the disk vertical structure.\\

ii) Dominance of P$_{rad}$, P$_{tot}$ = P$_{rad}$\\

 In this case, we cannot get a simple analytical conclusion, so for the radiation-dominated disk branch its vertical structure may - and likely will - be modified, as argued by \citet{2015begelman}, and the strength of this modification depends on specific parameters. 

\citet{2015begelman} derives a formula that parameterizes the strength of the effect as the ratio of the standard disk height plus the dead zone to the standard disk height (see their Equation 23). The formula has a strong and interesting dependence on the dimensionless accretion rate with the power 29/99. We thus apply it to modify the disk structure.

\subsection{Modification of the cold disk structure due to the magnetic field and stationary disk equations}
\label{sect:begelman}


In our model, we use a two-layer disk/corona approach as in \citet{2007janiuk} instead of a full vertical structure. Corona is unaffected but we incorporate the possible effective modification of the disk zone due to the action of the magnetic field. Equation 23 in \citet{2015begelman} implies the rise of the role of the magnetic field with dimensionless accretion rate. Since the application of the correction term to vertically-averaged disk structure is not unique, we introduce two specific prescriptions.\\

\begin{figure}
    \centering
     \includegraphics[scale=0.7]{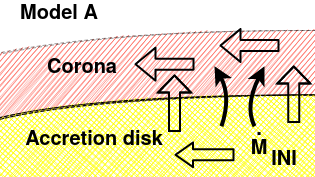}
    \includegraphics[scale=0.7]{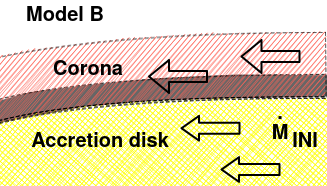}
    \caption{A zoomed-in schematic view of the accretion flow from Figure \ref{fig:our_model} with introduced modified structures in the disk. Model A (upper panel) shows energy transfer by the magnetic field in the form of Alfven waves represented as curved solid black arrows. Model B (bottom panel) shows the 'dead zone' as in the gray area between the corona and accretion disk.}
    
    \label{fig:zoom_modelab}
\end{figure}

(i) Model A \\ 

In this approach we follow the general idea presented in \citet{czerny2003} where we introduced energy transfer by the magnetic field in the form of Alfven waves (upper panel of Figure \ref{fig:zoom_modelab}). This energy flux was thus proportional to the magnetic field $B^2$, the Alfven speed, $v_A$, and the surface coverage by active regions. For this last effect, we did not have there a convincing parametrization (disk thickness was used). Now we propose to use the scaling with accretion rate from the \citet{2015begelman}. Thus the new energy flux becomes
\begin{equation}
    F_{mag} = b P v_A (\dot m)^{29/99},
    \label{eq:mag_flux_A}
\end{equation}
where the magnetic field is assumed to scale with $P_{tot}$, and $b$ is the arbitrary scaling constant. This magnetic energy transport is added to the general disk energy balance. Such a scaling implies that the role of the magnetic field is low at small accretion rates (when $P_{gas}$ dominates), and rises for higher accretion rates. Thus this formula is a smooth function with expected properties but without any onset treshold.\\

(ii)  Model B \\  

Following the idea of the 'dead zone' of \citet{2015begelman}, we can expect that the role of the magnetic field will scale with the relative width of the 'dead zone' (bottom panel of Figure \ref{fig:zoom_modelab}). Since the scaling in Equation (23) of \citet{2015begelman} give the ratio of $z_2/z_1$, where $z_2$ is the zone plus the disk thickness while $z_1$ is the disk thickness, the global effect should scale as $(z_2 - z_1)/z_2$. We cannot use the Compton coefficient here as in our time-dependent simulations of the mean zone such a factor does not play a role, so we use again the scaling  from Equation \ref{eq:mag_flux_A} thus arriving at the following modification term in our energy balance
\begin{equation}
    F_{mag} = F_{tot} b'(1 - \frac{F_{tot}}{P v_A (\dot m)^{29/99}}),
     \label{eq:mag_flux_B}
\end{equation}
where $F_{tot}$ is the total local dissipated flux. It requires the condition that the term $P v_A (\dot m)^{29/99}$ is larger than 1, otherwise, the term would be nominally negative, and we then set it to zero. Here $b'$ is again an arbitrary proportionality constant.

We combine this new energy flux term (in the form of Equation~\ref{eq:mag_flux_A} or Equation~\ref{eq:mag_flux_B})  with the usual equations of the stationary disk structure as follows.
We use the vertically averaged relations for the total energy flux dissipating in the accretion disk (Equation 19 in \citealt{2002janiuk})
\begin{equation}
    F_{tot} = C_1 \frac{3}{2} \alpha P \Omega_K H,
\end{equation}
which in this case is generated through the $\alpha$ viscosity, where 
\begin{equation}
    C_1 = \frac{ \int_{0}^{H} P dz }{P_e H}
\end{equation}
the local emitted flux
\begin{equation}
    F_{em} = C_2 \frac{ac T_e^4}{3 \kappa \rho_e H}, 
\end{equation}
\begin{equation}
    C_2 = \frac{\rho_e F_{rad}H}{\int_{0}^{H} \rho F dz}
\end{equation}
and the advection component

\begin{equation}
    F_{adv} = \frac{1}{2\pi r^2} \frac{\dot M P q_{adv}}{\rho}.
\end{equation}
Here the values $T_e$, $\rho_e$, $P_e$ are the equatorial values of the temperature, density, and pressure, correspondingly, and $H$ is the disk thickness. The coefficients $C_1$ and $C_2$ were calculated from the disk vertical structure at a fixed radius and then used as universal parameters \citep[see][]{2002janiuk}. Combining these equations, we could write the energy balance equation for a given radius
\begin{equation}
   C_1 \frac{3}{2} \alpha P_e \Omega_K H = C_2 \frac{ac T_e^4}{3 \kappa \rho H} -  \frac{1}{2\pi r^2} \frac{\dot M P_e q_{adv}}{\rho}
    \label{eq:energybalanceold}
\end{equation}
as was presented in \citep{2002janiuk}. Here, for the simplicity of the notation, we already dropped the indices $e$ marking equatorial plane values.

Finally, by adding the extra magnetic transport cooling component to the energy balance equation (Equation \ref{eq:energybalanceold}) we obtain:

\begin{equation}
   C_1 \frac{3}{2} \alpha P \Omega_K H = C_2 \frac{ac T^4}{3 \kappa \rho H} -  \frac{1}{2\pi r^2} \frac{\dot M P q_{adv}}{\rho} -  F_{mag},
    \label{eq:energybalancenew}
\end{equation}
with the prescription for $F_{mag}$ following Equation~\ref{eq:mag_flux_A} or \ref{eq:mag_flux_B}.

Here we assumed $\alpha P_{tot}$ viscosity, but we also consider $\alpha \sqrt{P_{tot}P_{gas}}$ assumption \citep[see][]{szuszkiewicz1990,janiuk2011,grzedzielski2017} which requires replacement of the corresponding dissipation term in the equations above. In the list of models, we refer to the first option as {\it iPtot}, and to the second option as {\it isqrt}.

\begin{figure*}[ht]
    \centering
    \includegraphics[width=6cm]{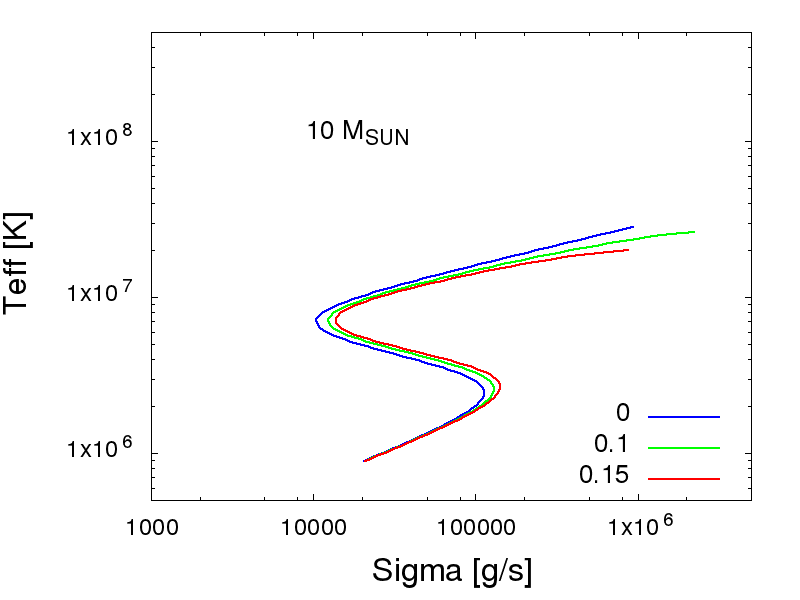}
     \includegraphics[width=6cm]{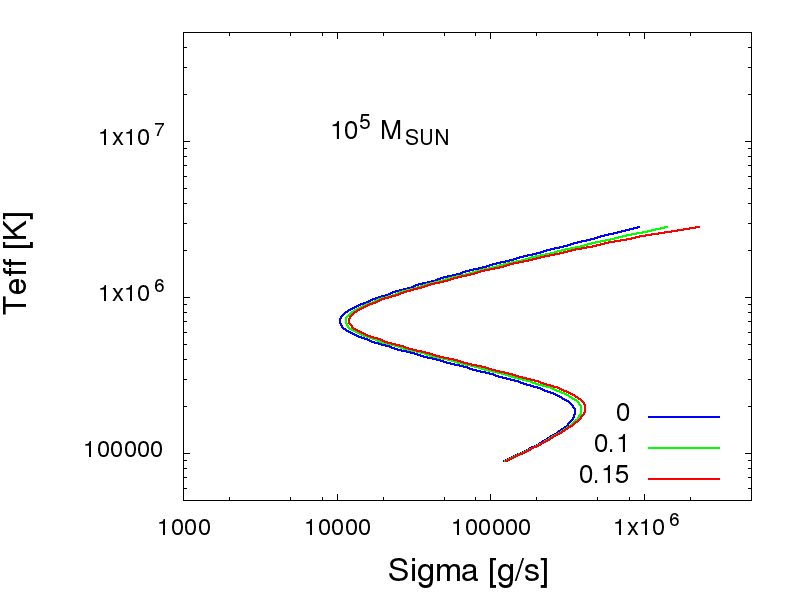}
     \includegraphics[width=6cm]{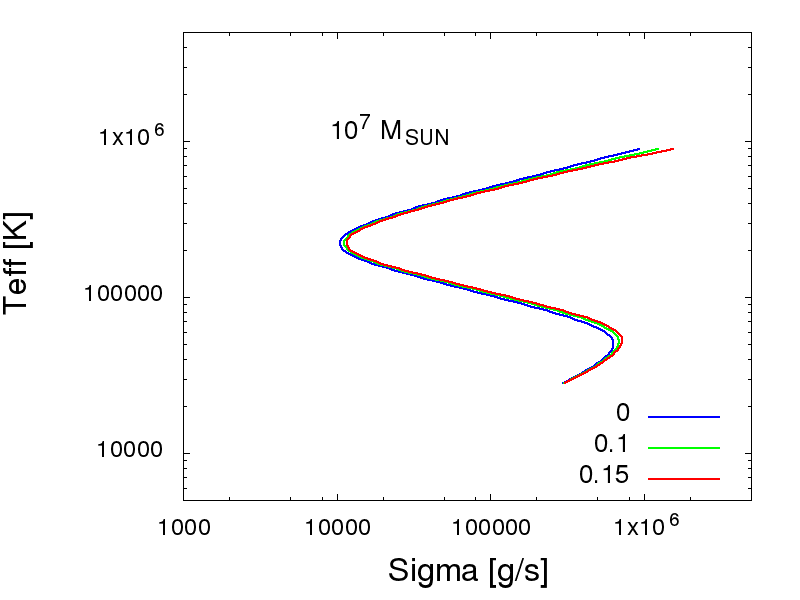}
    \caption{Local stability curves for black hole masses  for 10M$\odot$ (left panel), 10$^5$M$\odot$ (middle panel) and 10$^7$M$\odot$ (right panel). For each curve we keep R=10R$_{schw}$ and $\alpha$ = 0.01. The color of the S-curve represents different coefficient value for Model A: blue for  $b$ = 0 (base model), green for $b$ = 0.1, and red for $b$ = 0.15.  For each plot we keep the same X-axis and the same range (3 orders of magnitude) for Y-axis.}
    \label{fig:mnew2d-czlon}
\end{figure*}

These equations allow to obtain the local stability curves for stationary models at each radius, set by the value of the black hole mass, viscosity parameter $\alpha$ and the magnetic field parameter $b$ or $b'$. 


\begin{figure*}[ht]
    \centering
    \includegraphics[width=6cm]{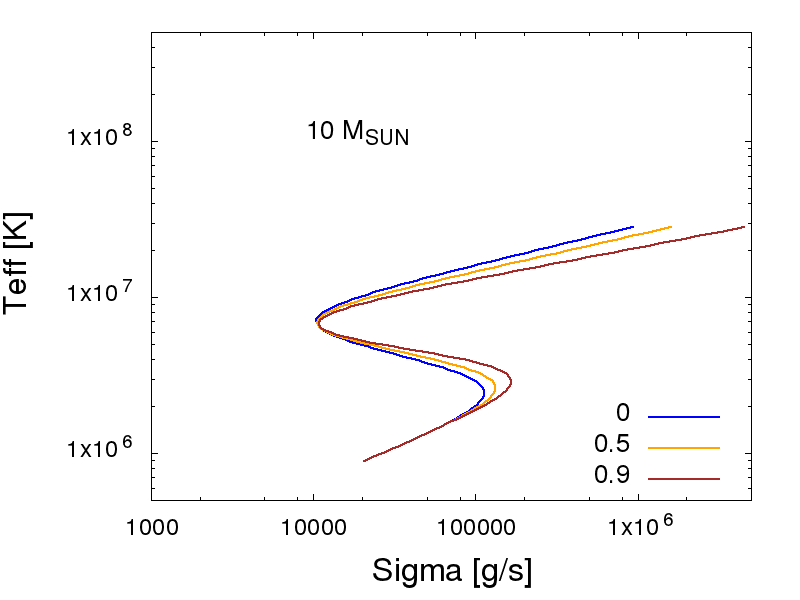}
     \includegraphics[width=6cm]{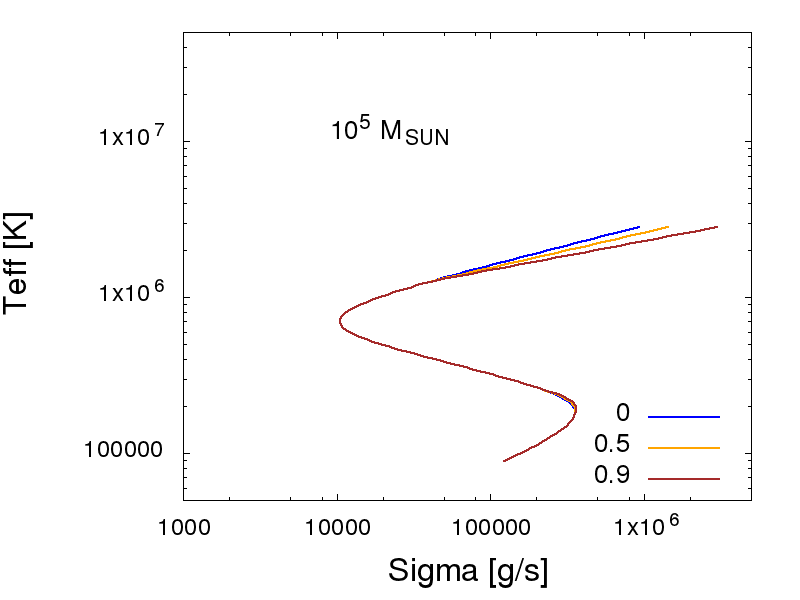}
     \includegraphics[width=6cm]{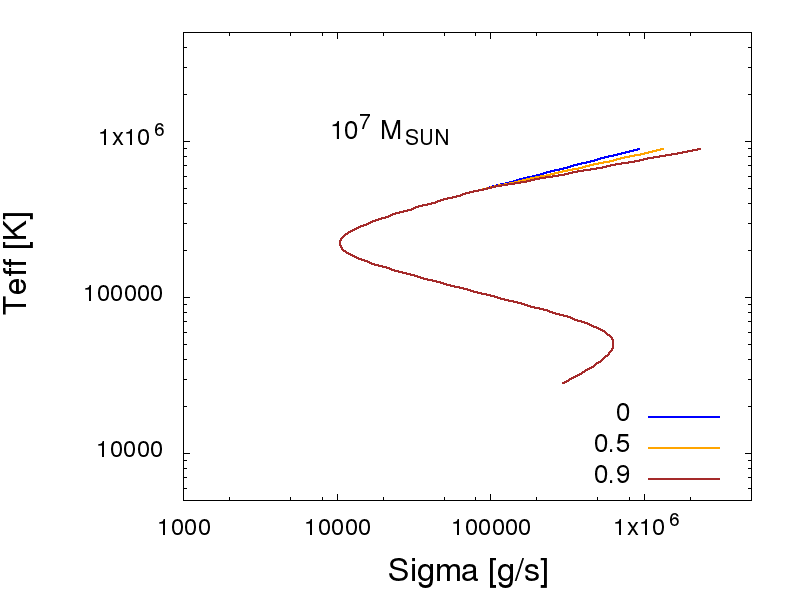}
    \caption{Local stability curves for black hole masses  for 10M$\odot$ (left panel), 10$^5$M$\odot$ (middle panel) and 10$^7$M$\odot$ (right panel). For each curve we keep R=10R$_{schw}$ and $\alpha$ = 0.01. The color of the S-curve represents different coefficient value for Model B: blue for $b'$ = 0 (base model), orange for $b'$ = 0.5, and brown for $b'$ = 0.9.  For each plot we keep the same X-axis and the same range (3 orders of magnitude) for Y-axis.}
    \label{fig:mnew2d-1przezczlon}
\end{figure*}



\subsection{Time evolution of the disk/corona system}

Apart from the modifications described in Sect.~\ref{sect:ADAF} and \ref{sect:begelman}, we follow the time evolution of the disk/corona system as described in \citet{2007janiuk}. Both the disk and the corona evolve, and the disk mass is evaporated slowly to the corona according to Equation~10 of \citet{2007janiuk}. The local effect of the magnetic field related to MRI is included assuming that its time-dependence can be well represented by the Markoff process. The stochastic variations in the magnetic field affect the coronal outflow, and, indirectly, the underlying disk. This model thus includes the stochastic variability of the flow which may be coupled with large-scale radiation pressure instability leading to periodic outbursts. Since we now introduce additional modifications concerning the model of \citet{2007janiuk}, we give below the full set of time-dependent equations.

For the accretion disk evolution, we solve two equations. The equation of mass and
angular momentum conservation
\begin{eqnarray}
{\partial \Sigma \over \partial t}={1 \over r}{\partial \over \partial
r}(3 r^{1/2} {\partial \over \partial r}(r^{1/2} \nu \Sigma))
- \dot m_{\rm z}, 
\end{eqnarray}

and the energy equation

\begin{eqnarray}
{\partial T \over \partial t} + v_{\rm r}{\partial T \over \partial r}
= {T \over \Sigma}{4-3\beta \over 12-10.5\beta}
({\partial \Sigma \over
\partial t}+  v_{\rm r}{\partial \Sigma \over \partial r}) 
\nonumber
\\
+{T\over P H}{1\over 12-10.5\beta} 
(Q^{+}-Q^{-}-\ F_{mag}),
\end{eqnarray}


In the case of the equation of mass and angular momentum conservation, we use the following term (one from three possible mechanisms, see \cite{2007janiuk}, and their Equation 10 (case B) for the disk evaporation $\dot m_{\rm z}$ into corona: 
 ${B^2 \over 8 \pi} c {\mu m_p \over k T_{vir}} $). The magnetic field in this model evolves in a complex way: fast intrinsic variation is modelled as a stochastic variability related to the local MRI and evolved as a discreet Markoff process as suggested by \citet{king2004} and \citet{mayer2006}, and part of the evolution is coupled to the change of the interior disk parameters since the maximum of the magnetic field is set by the value of the total pressure at the equator, i.e.
 \begin{equation}
 {B^2 \over 4 \pi} = u_n^2 \alpha P.
 \end{equation}
 Here the Markoff chain value $u_n$ has the mean value zero, the dispersion 1, the memory coefficient -0.5, and the local time step is proportional to the local dynamical timescale (see \citealt{2007janiuk}).
 
In the case of the energy equation 
$\beta$ is $P_{gas} \over P_{tot}$, and $v_{\rm r}$ is the radial velocity in the disk, 
Q$^{+}$ is viscous heating, Q$^{-}$ is radiative cooling, and 
F$_{mag}$ is an extra cooling component (from Model A or Model B), with dimensionless factor, $b$, which possible values we investigate in this work. 

For the evolution of corona, we solve only the mass and angular momentum conservation, since we assume the virial temperature so the temperature at each radius is fixed.

\begin{eqnarray}
{\partial \Sigma_{\rm cor} \over \partial t}={1 \over r}{\partial \over \partial
r}(3 r^{1/2} {\partial \over \partial r}(r^{1/2} \nu_{\rm cor} \Sigma_{\rm cor}))
+ \dot m_{\rm z}.
\label{eq:sigcorevol}
\end{eqnarray}

\section{Results}

Before we show the results for the time evolution of the disk under the radiation pressure instability, we shortly present a few characteristic properties of the current model, modified concerning the one used by \citet{2007janiuk}. This helps later to understand some of the new evolutionary trends. Since stationary models are purely local (interaction with nearby radii are set by the stationarity condition), they can only display well the role of the magnetic field through modification of the disk vertical structure. 

\subsection{Stationary solutions and exemplary stability curves}

\label{s_curves}

As we show in Sec. \ref{sect:begelman}, the modification of the energy flux by mimicking a magnetic field influences the energy balance (Equation \ref{eq:energybalancenew}). 
We show the local surface density vs. temperature plane 'S-curve' for Model A for various coefficients in Figure \ref{fig:mnew2d-czlon} and Model B for various coefficients in Figure \ref{fig:mnew2d-1przezczlon}. These curves represent the local disk structure under the condition of thermal balance. Since those are stationary models, various valued of the disk temperature correspond to various values of accretion rate. We plot these for 3 different values of black hole masses: 10  $M_\odot$, 10$^5 M_\odot$ and 10$^7 M_\odot$. In all cases, we performed computations at the radius of 10$R_{Schw}$ and we assume the viscosity parameter $\alpha$ = 0.01.

The shape of the 'S-curve' provides direct information about the local stability of the disk since positive branches correspond to stable solutions while negative slopes correspond to unstable ones. The disk evolution estimated locally would manifest as a loop on these plots, with the stages of slow evolution along the stable branches and rapid vertical transitions in thermal timescales between the stable branches. The extension of the unstable branch allows for an estimate of the amplitude and the duration of the full cycle but only locally. Global evolution is a combined effect of the processes at all radii. Nevertheless, the change in the 'S-curve' shape with an introduction of the magnetic field effects shows the likely trend in the global evolution. 

In the case of Model A, S-curves are shifted horizontally (besides of lower stable branch) and the lower turning point seems to slightly shift vertically. The effect vanishes with increasing the black hole mass. The separation between the minimum and the maximum value of the temperature at the unstable branch becomes somewhat lower with the rising value of the parameter $b$ so we rather expect lower outburst amplitude. The minimum value of the density at the unstable branch becomes higher, so after the collapse of the disk to the lower branch its local accretion rate will be higher and viscous timescales shorter with the rise of $b$. The effect of the magnetic field, however, seems to be relatively weaker for larger masses.

In the case of  Model B, the flattening of the unstable part of the curve again is observed, implying expectations of a lower outburst amplitude. However, the upper point of the unstable branch seems to be unaffected by the change of the parameter $b'$, so the outburst timescale may not be reduced in this case. Again, the changes in the 'S-curve' shape at higher accretion rates are smaller.
A similar effect (in the shape) of the upper branch of stability curves was obtained by introducing vertical outflows \cite[][see Figure 3 therein]{2002janiuk}.

Summarising, the presented curves are modified by the effect of the magnetic field in the disk interior but modifications in both models (A and B) do not seem to be very strong, especially in the case of higher black hole masses.
However, the local study of the 'S-curve' can only indicate trends while actual properties of the disk time evolution can be only assessed by performing global evolution of the disk. The effects of the inner and outer boundary conditions are seen only in the global simulations.


\subsection{Time-dependent evolution}


\begin{figure*}[ht]
    \centering
      \includegraphics[width=8cm]{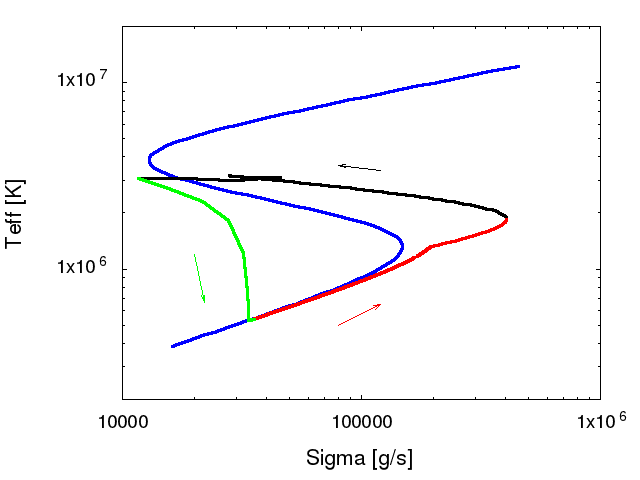}
       \includegraphics[width=8cm]{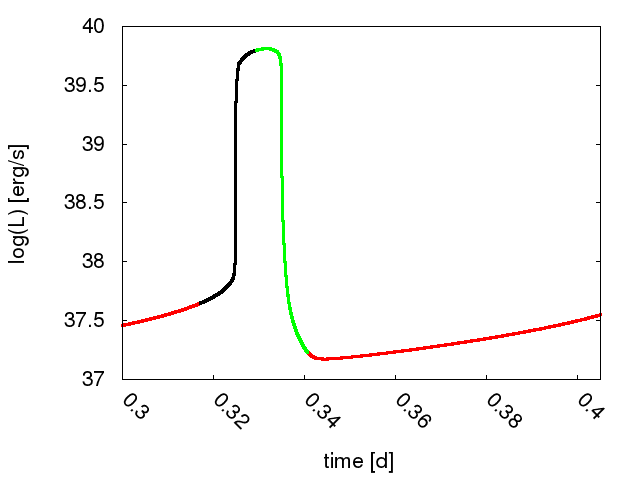}
    \caption{Left panel: Local stability curve (blue line) with marked outburst cycle for black hole mass 10M$\odot$ and R=40R$_{schw}$ for base model without R$_{ADAF}$ and  R$_{out}$ = 300R$_{schw}$.   Right panel: the exemplary fragment of the disk lightcurve with cycle-steps marked for black hole mass 10M$\odot$. In colors we marked the following steps of the cycle: the black part of the light curve represents the heating phase, the green one the advective phase, and the red one the diffusive phase. The colored arrows mark the direction of the cycle - anticlockwise.}
    \label{fig:basemodel-scurves-sigte}
\end{figure*}

Our global model of the accretion disk evolution has the following input parameters: black hole mass, $M_{BH}$, accretion rate $\dot m$ in Eddington units, the inner radius $R_{ADAF}$ (or no ADAF and then the inner radius is located at ISCO), outer radius $R_{out}$, type of the viscosity law (viscosity parameter $\alpha$ is always fixed at the same value, 0.01), type of the magnetic field modification and the coefficient ($b$ or $b'$) scaling the strength of this effect, and the presence or the absence of the hot accreting corona above the disk. We calculated several such global models, and their input parameters are listed in Tables~\ref{tab:modele}, \ref{tab:modele-10-5-xicor-0}, and \ref{tab:modele-10-7-xicor-0}. 

For each of the models, we calculate the standard output parameters: the period (if outbursts are present) which is measured from the peak to the next peak, the relative amplitude of the disk outburst (maximum to minimum flux), and the relative outburst amplitude of the corona. These results are given in the corresponding tables. Outburst amplitudes in the disk correspond to the total time-dependent luminosity of the disk, $F_{tot}$, coming from the local radiative flux integrated over the disk surface for each moment of the time evolution. The corona luminosity is calculated in a time-dependent manner, first locally as the dissipation rate in accreting coronal flow, assuming $\alpha$ viscosity as in the disk, and then by integrating the coronal flux over the disk surface.  We illustrate the results with exemplary lightcurves.


From the point of view of the evolutionary timescales, the black hole mass is the key parameter as it determines the object's scale size. So we first concentrate on the results for the $M_{BH} = 10 M_{\odot}$ and later proceed toward larger masses. 

Before plotting the full lightcurves we first illustrate an important difference between the stability curves discussed in Section~\ref{s_curves} and the full time-dependent computations. For that we selected a model with $M_{BH} = 10 M_{\odot}$, accretion rate $\dot m = 0.67$, without a hot corona, or inner ADAF, and with the outer radius $R_{out} = 300 R_{schw}$, large enough to cover the whole instability region, and without any extra effects of the magnetic field. We select the representative radius of $40 R_{Schw}$ and we plot in Figure~\ref{fig:basemodel-scurves-sigte}, left panel, the evolution of the effective temperature and the surface density at this radius. In the right panel, we plot the small fragment of the total disk lightcurve. We mark characteristic parts of the lightcurve with different colors for slow luminosity rise in viscous timescale in red, fast thermal rise, and relatively fast viscous evolution along the hot branch in black, and then the decay in green. The same colors are used in the left panel, with arrows showing the direction of the limit cycle. As a reference, we plot the stationary stability S-curve for their model at this radius, but we see that actual evolution does not follow the stationary curve perfectly, since the disk evolved also at smaller and larger radii, and radial derivatives which are affecting the local disk state are also time-dependent.

\subsubsection{Microquasars}
\label{sect:micro}

In this subsection, we fix the black hole mass value at $10 M_{\odot}$ so we concentrate on solutions eventually applicable to microquasars showing heart-beat states. 

We calculated a large set of models without the inner ADAF (i.e. $R_{ADAF} = 3 R_{Schw}$), with the large outer radius not affecting the disk evolution ($R_{out} = 300 R_{Schw}$, the disk is already stable there), but without or with the corona, and with increasing magnetic field parameter. For each of these models, we used the viscous torque proportional to the total (i.e. gas + radiation) pressure. A test model, with the square root law for the viscous torque, was stable. As we see from Table~\ref{tab:modele}, the presence of the corona does not affect considerably the amplitude and the timescales of the outbursts, so we plot the solutions with corona, as they contain extra information on the expected behaviour of the hard X-rays.

\begin{figure*}[ht]
    \centering
      \includegraphics[scale=0.35]{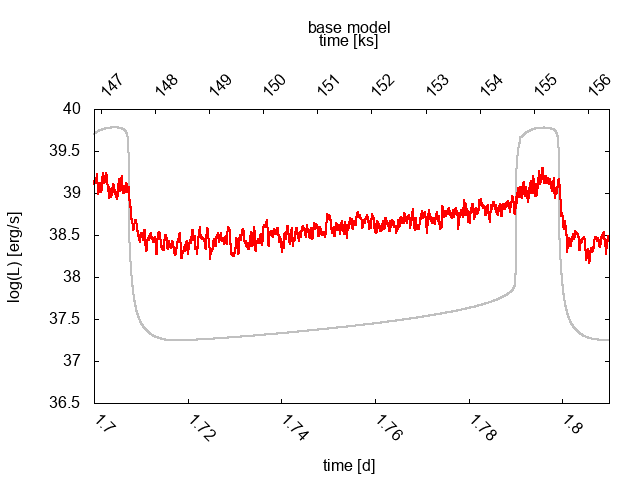}
       \includegraphics[scale=0.35]{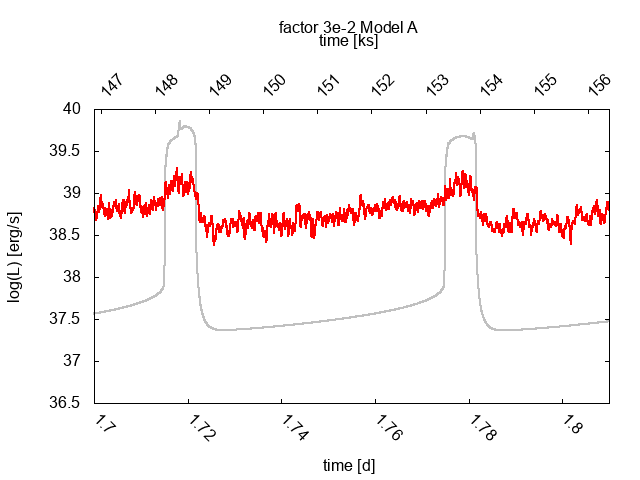}
              \includegraphics[scale=0.35]{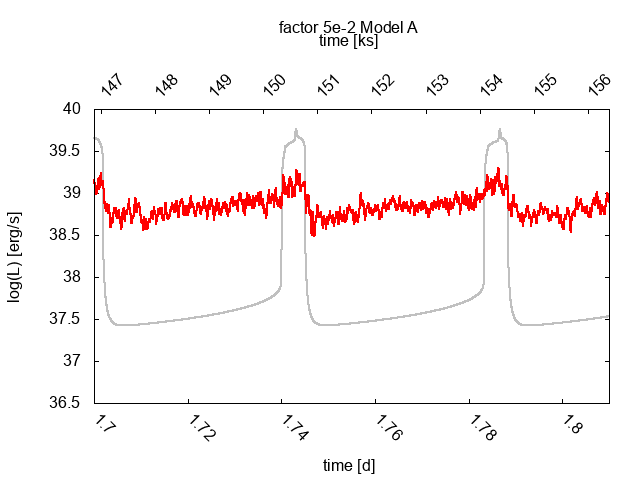}
       \includegraphics[scale=0.35]{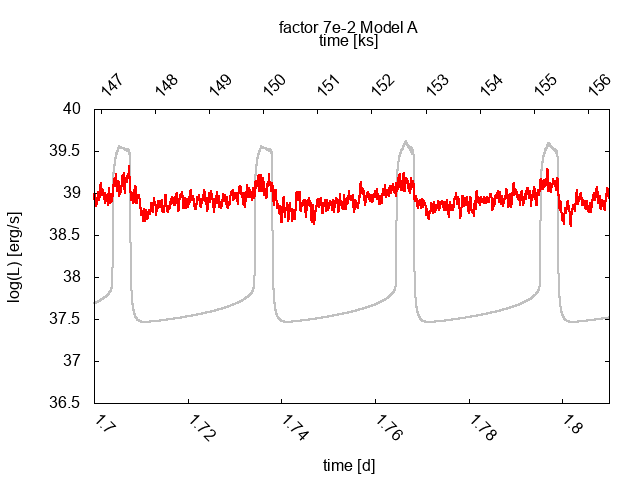}
          \includegraphics[scale=0.35]{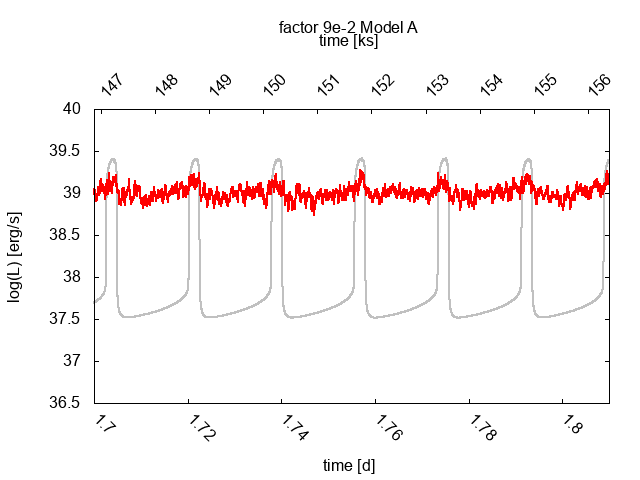}
        \includegraphics[scale=0.35]{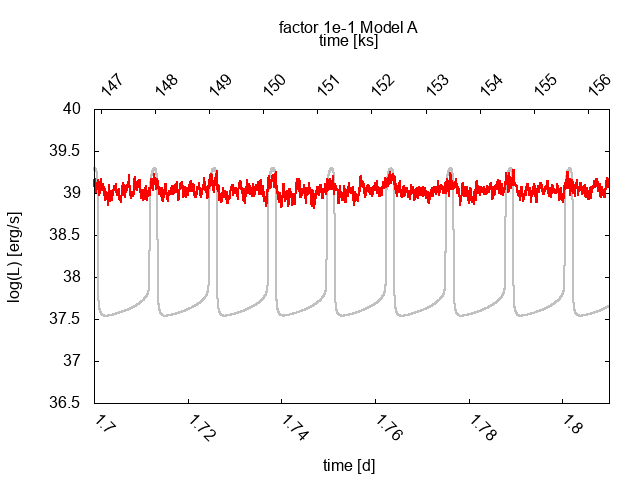}
   \includegraphics[scale=0.35]{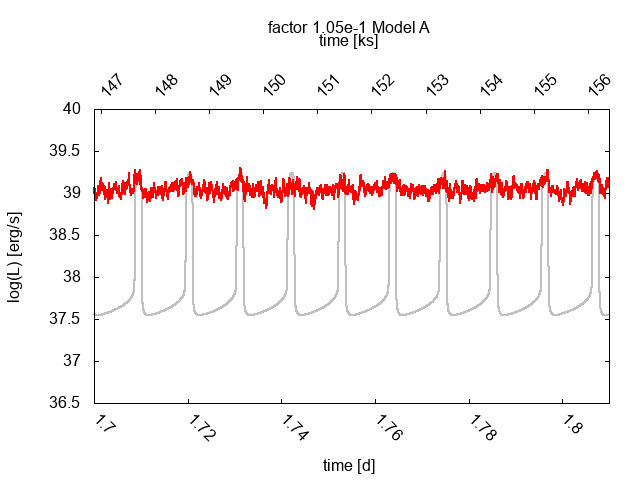}
   \includegraphics[scale=0.35]{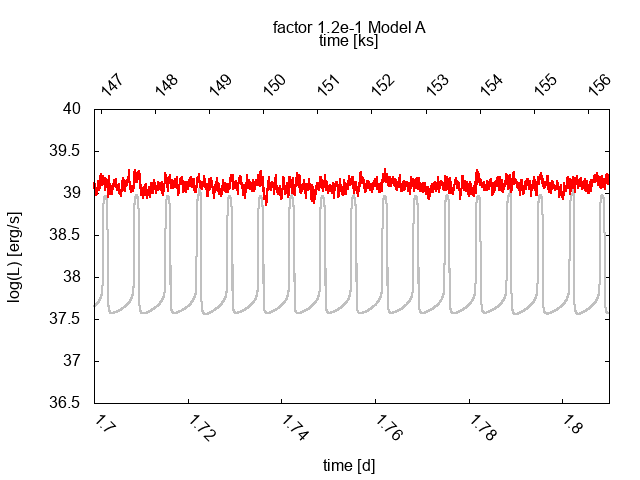}      
    \includegraphics[scale=0.35]{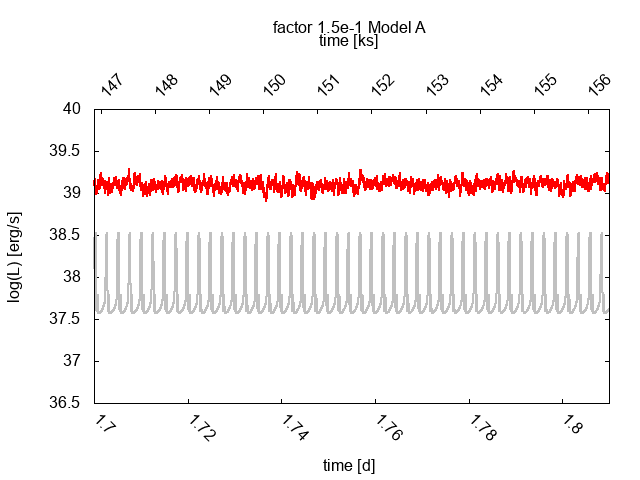}
             \caption{The disk (marked in gray) and corona (marked in red) lightcurves for 10M$\odot$
       for different coefficient value $b'$ for Model A (from the top left): $b$ = 0 (base model),   0.01,  0.03,  0.05, 0.07, 0.09, 0.1, 0.105, 0.12 and 0.15. For all cases the inner radius is 3 R$_{schw}$ and the outer radius R$_{out}$ = 300  R$_{schw}$.}
             
    \label{fig:mqso-factor-modela-corona-xi0}
\end{figure*}

We first analyse model A of the magnetic field effect. In Figure~\ref{fig:mqso-factor-modela-corona-xi0} we plot the lightcurves for the disk and for the corona using the same timescale in the horizontal axis.
From this sequence of models we see that the outburst period is strongly reduced with the rise of the parameter $b$ (see Equation ~\ref{eq:mag_flux_A}). Without the role of the magnetic field $b = 0$, the outburst period is about 0.1 day (2.4 hours), while at the value $b = 0.15$ the period is reduced to 2 minutes. Higher values of the parameter $b$ give stable solutions. The reduction of the period is also accompanied by the reduction of the outburst amplitude. It again illustrates the fact that stationary stability curves are shown in Figure~\ref{fig:mnew2d-czlon} and \ref{fig:mnew2d-1przezczlon} do not predict well the outbursts, and computations of actual time-dependent global solutions are necessary. The outburst amplitudes in the corona are much smaller than the disk amplitudes in our model, and at $b > 0.1$ the corona is almost stable apart from the usual stochastic variability. In observations the outbursts amplitudes are measured using countrates in the selected energy band while in our theoretical model we calculate bolometric luminosities., therefore the direct comparison with the data is not simple, but the radiation pressure instability model well represented the observed time delays between the hard X-ray and soft X-ray flux \citep{2005janiuk}. Here we do not calculate the time delay as we rather concentrate on amplitudes and timescales.

\clearpage
\onecolumn
\begin{landscape}
\begin{longtable}{c|c|c|c|c|c|c|c|c|c|c}
\caption{The list of models calculated in the present paper for 10 $M_{\odot}$.} \label{tab:modele}
\\
$M_{BH}$ & $\dot m$ & \textbf{ $R_{ADAF}$} & \textbf{$R_{out}$} & corona & viscosity & magnetic & magnetic & period& amplitude&amplitude\\ 
$[M_{\odot}]$ & [$\dot M_{Edd}$] & [$R_{Schw}$] & [$R_{Schw}$] &  & type &  & coefficient & & disk&corona\\ \hline

\endhead

10    &  0.67  & -  &  - & - & isqrt & - & - & stable & - & - \\
10    &  0.67  & -  &  - & - & iPtot & - & - &0.1 [d] & 10$^{2.63}$& -\\
10    &  0.67  & -  &  - & yes & iPtot & - & - & 0.09 [d] & 10$^{2.51}$& 10$^{0.79}$\\ 
10    &  0.67  & -  &  - & - & iPtot & Model A & 1e-2 & 0.079 [d] & 10$^{2.49}$& -\\
10    &  0.67  & -  &  - & - & iPtot & Model A & 3e-2& 0.059 [d] & 10$^{2.35 }$& -\\
10    &  0.67  & -  &  - & - & iPtot & Model A & 5e-2& 0.042 [d] & 10$^{2.23}$& -\\
10    &  0.67  & -  &  - & - & iPtot & Model A & 7e-2& 0.028 [d] & 10$^{2.097}$& -\\
10    &  0.67  & -  &  - & - & iPtot & Model A & 9e-2& 0.017 [d] & 10$^{1.98}$& -\\
10    &  0.67  & -  &  - & - & iPtot & Model A & 1e-1& 0.011 [d] & 10$^{1.68}$& -\\
10    &  0.67  & -  &  - & - & iPtot & Model A & 1.05e-1& 0.009 [d] & 10$^{1.59}$& -\\
10    &  0.67  & -  &  - & - & iPtot & Model A & 1.2e-1& 0.0046 [d] & 10$^{1.306}$& -\\
10    &  0.67  & -  &  - & - & iPtot & Model A & 1.5e-1 & 0.00152 [d] & 10$^{1.01}$& -\\
10    &  0.67  & -  &  - & - & iPtot & Model A & 2e-1 & stable& -&-\\
10    &  0.67  & -  &  - & yes & iPtot & Model A & 1e-2 &0.079 [d] & 10$^{2.403}$& 10$^{0.64}$\\
10    &  0.67  & -  &  - & yes & iPtot & Model A & 3e-2  & 0.059 [d] & 10$^{2.403}$& 10$^{0.54}$\\
10    &  0.67  & -  &  - & yes& iPtot & Model A & 5e-2  & 0.042 [d] & 10$^{2.193}$& 10$^{0.359}$\\
10    &  0.67  & -  &  - & yes & iPtot & Model A  & 7e-2  & 0.0314 [d] & 10$^{2.07}$& 10$^{0.318}$\\
10    &  0.67  & -  &  - & yes& iPtot & Model A & 9e-2 & 0.0175[d] & 10$^{1.882}$& 10$^{0.2}$\\
10    &  0.67  & -  &  - & yes & iPtot & Model A & 1e-1  & 0.01 [d] & 10$^{1.74}$& 10$^{0.17}$\\
10    &  0.67  & -  &  - & yes & iPtot & Model A & 1.05e-1 & 0.01 [d] & 10$^{1.683}$& 10$^{0.1}$\\
10    &  0.67  & -  &  - & yes & iPtot & Model A & 1.2e-1 & 0.006 [d] & 10$^{1.374}$& stable\\
10    &  0.67  & -  &  - & yes & iPtot & Model A & 1.5e-1 & 0.0024[d] & 10$^{0.94}$& stable\\
10    &  0.67  & -  &  - & yes & iPtot & Model A & 2e-1& stable&-&-\\

10    &  0.67  & -  &  - & - & iPtot & Model B &  5e-2&0.089[d] & 10$^{2.59}$& -\\ 
10    &  0.67  & -  &  - & - & iPtot & Model B &  1e-1 & 0.081[d] & 10$^{2.53}$& -\\
10    &  0.67  & -  &  - & - & iPtot & Model B &  2e-1 & 0.073[d] & 10$^{2.54}$& -\\ 
10    &  0.67  & -  &  - & - & iPtot & Model B &  5e-1& 0.052[d] & 10$^{2.32}$& -\\ 
10    &  0.67  & -  &  - & - & iPtot & Model B &  7e-1& 0.035[d] & 10$^{2.19}$& -\\ 
10    &  0.67  & -  &  - & - & iPtot & Model B &  9e-1& 0.019[d] & 10$^{2.08}$& -\\ 
10    &  0.67  & -  &  - & - & iPtot & Model B &  9.5e-1&  0.012[d] & 10$^{1.86}$& -\\ 
10    &  0.67  & -  &  - & - & iPtot & Model B &  9.7e-1&  0.012[d] & 10$^{1.79}$& -\\ 
10    &  0.67  & -  &  - & - & iPtot & Model B &  1.0 & 0.0088[d] & 10$^{1.62}$& -\\ 
10    &  0.67  & -  &  - & - & iPtot & Model B &  1.1 & stable&-&-\\ 
10    &  0.67  & -  &  - & - & iPtot & Model B &  1.3 & stable&-&-\\ 
10    &  0.67  & -  &  - & - & iPtot & Model B &  1.5 & stable&-&-\\ 
10    &  0.67  & -  &  - & yes & iPtot & Model B &  5e-2  & 0.09[d] & 10$^{2.49}$& 10$^{0.69}$\\ 
10    &  0.67  & -  &  - & yes & iPtot & Model B &  1e-1  &0.08[d] & 10$^{2.46}$& 10$^{0.66}$\\ 
10    &  0.67  & -  &  - & yes & iPtot & Model B &  2e-1  &0.072[d] & 10$^{2.38}$& 10$^{0.64}$\\ 
10    &  0.67  & -  &  - & yes & iPtot & Model B &  5e-1& 0.046[d] & 10$^{2.25}$&  10$^{0.46}$\\ 
10    &  0.67  & -  &  - & yes & iPtot & Model B &  7e-1 & 0.029[d] & 10$^{2.06}$&  10$^{0.31}$\\ 
10    &  0.67  & -  &  - & yes & iPtot & Model B &  9e-1 & 0.014[d] & 10$^{1.75}$&  stable\\ 
10    &  0.67  & -  &  - & yes & iPtot & Model B &  9.5e-1& 0.012[d] & 10$^{1.57}$&  stable\\ 
10    &  0.67  & -  &  - & yes & iPtot & Model B &  9.7e-1& 0.011[d] & 10$^{1.49}$&  stable\\ 
10    &  0.67  & -  &  - & yes & iPtot & Model B &  1.0 & 0.027[d] & 10$^{1.34}$&  stable\\ 
10    &  0.67  & -  &  - & yes & iPtot & Model B &  1.1 & stable&-&-\\ 
10    &  0.67  & -  &  - & yes & iPtot & Model B &  1.3 &stable&-&- \\
10    &  0.67  & -  &  - & yes & iPtot & Model B &  1.5 &stable&-&- \\
10    &  0.67  & -  &  - & yes & iPtot & Model B & 2e-1&stable&-&- \\
10    &  0.67  & 5  &  - & yes & iPtot & - & - & 0.10[d] & 10$^{2.44}$&10$^{0.68}$ \\
10    &  0.67  & 7  &  - & yes & iPtot & - & - & 0.124[d] & 10$^{2.56}$&10$^{0.73}$ \\
10    &  0.67  & 10  &  - & yes & iPtot & - & -& 0.096[d] & 10$^{2.14}$&10$^{0.758}$ \\
10    &  0.67  & 20  &  - & yes & iPtot & - & -& 0.11[d] & 10$^{1.57}$&10$^{0.47}$ \\
10    &  0.5  &   -&  - &-  & iPtot & Model B & 0.01 & 0.06 [d] & 10$^{2.43}$&-\\
10    &  0.5  &   -&  - & - & iPtot & Model B & 0.05 & 0.04 [d] & 10$^{2.23}$&-\\
10    &  0.5  &  - &  - &  -& iPtot & Model B & 0.07 & 0.03 [d] & 10$^{2.12}$&-\\ \hline \hline

\end{longtable}
\footnotesize{{\sc Notes.} Columns are as follows: (1) Black hole mass (2) accretion rate (3)  the inner ADAF location (4) the outer radius location (5) presence of corona (6) viscosity type (7) magnetic field model used in the simulation (see Sec. \ref{sec:modela_modelb}), (8)  magnetic field coefficient used in simulation, (9) period of the outbursts, (10) amplitude of the disk outburst, and (11) amplitude of the corona outburst.}
\end{landscape}
\clearpage
\twocolumn


\clearpage
\onecolumn
\begin{landscape}
\begin{longtable}{c|c|c|c|c|c|c|c|c|c|c}
\caption{The list of models calculated in the present paper for 10$^5$ $M_{\odot}$.}     \label{tab:modele-10-5-xicor-0}
\\
$M_{BH}$ & $\dot m$ & \textbf{ $R_{ADAF}$} & \textbf{$R_{out}$} & corona & viscosity & magnetic & magnetic & period& amplitude&amplitude\\ 
$[M_{\odot}]$ & [$\dot M_{Edd}$] & [$R_{Schw}$] & [$R_{Schw}$] &  & type &  & coefficient & & disk&corona\\ \hline

\endhead
$10^5$    &  0.50  & -  &  300 & - & iPtot &  -&- & 26.834[y] &10$^{4.392}$ &-\\
$10^5$    &  0.50  & -  &  50 & - & iPtot &  -&- &  0.89[y] & 10$^{3.78}$ &-\\
$10^5$    &  0.50  & -  &  50* & - & iPtot &  -&- & 0.37[y] & 10$^{3}$ &-\\
$10^5$    &  0.50  & -  &  100 & - & iPtot &  -&- & 1.55[y]& 10$^{4.42}$ &-\\
$10^5$    &  0.50  & -  &  100* & - & iPtot &  -&- & 1.52[y] & 10$^{3.5}$ &-\\
$10^5$    &  0.50  & -  &  300 & yes & iPtot &  -&- & 33.54[y] &10$^{4.407}$ &10$^{2.68}$\\
$10^5$    &  0.50  & -  &  50 & yes & iPtot &  -&-&0.75[y]&10$^{3.29}$ & \\
$10^5$    &  0.50  & -  &  50* & yes & iPtot &  -&- & 0.42[y] &10$^{3.8}$ &10$^{2}$\\
$10^5$    &  0.50  & -  &  100 & yes & iPtot &  -&- & 1.55[y] &10$^{4.3}$ &10$^{2.25}$\\
$10^5$    &  0.50  & -  &  100* & yes & iPtot &  -&- & 1.51[y] &10$^{3.9}$ &10$^{2.2}$\\
$10^5$    &  0.50  & 5  &  300 & yes & iPtot &  -& -& 33.183[y] &10$^{4.2462}$ &10$^{2.6077}$\\
$10^5$    &  0.50  & 7  &  300 & yes & iPtot &  -& -& 32.84[y] &10$^{4.14}$ &10$^{2.41}$\\
$10^5$    &  0.50  & 10  &  300 & yes & iPtot &-  & -& 32.65[y] &10$^{4.02}$ &10$^{2.437}$\\
$10^5$    &  0.50  & 20  &  300 & yes & iPtot &-  & -& 31.95[y] &10$^{3.82}$ &10$^{1.95}$\\
$10^5$    &  0.50  & -  &  300 & - & isqrt &  -&  -&  stable&-&-\\
$10^5$     &  0.50  & -  &  - & - & iPtot & Model A & 1e-2 & 27.61[y] &10$^{4.33}$ &-\\
$10^5$     &  0.50  & -  &  - & - & iPtot & Model A & 3e-2 &31.07[y] &10$^{4.27}$ &-\\
$10^5$     &  0.50  & -  &  - & - & iPtot & Model A & 5e-2 & 34.28[y] &10$^{4.07}$ &-\\
$10^5$     &  0.50  & -  &  - & - & iPtot & Model A & 7e-2 &34.59[y] &10$^{3.73}$ &-\\
$10^5$     &  0.50  & -  &  - & - & iPtot & Model A & 9e-2 &16.29[y] &10$^{2.8}$ &-\\
$10^5$     &  0.50  & -  &  - & - & iPtot & Model A & 1e-1 &10.8[y] &10$^{2.66}$ &-\\
$10^5$     &  0.50  & -  &  - & - & iPtot & Model A & 1.2e-1 &4.78[y] &10$^{2.45}$ &-\\
$10^5$     &  0.50  & -  &  - & - & iPtot & Model A & 1.5e-1 &3.04[y] &10$^{2.11}$ &-\\
$10^5$     &  0.50  & -  &  - & - & iPtot & Model A & 1.7e-1 &2.33[y] &10$^{1.97}$ &-\\
$10^5$     &  0.50  & -  &  - & - & iPtot & Model A & 2e-1 & 1.35[y] &10$^{1.66}$ &-\\
$10^5$     &  0.50  & -  &  - & - & iPtot & Model A & 2.2e-1 &1.10[y] &10$^{1.29}$ &-\\
$10^5$    &  0.50  & -  &  - & - & iPtot & Model A &  3e-1 &stable&-&-\\
$10^5$     &  0.50  & -  &  - & yes & iPtot & Model A & 1e-2 & 35.7[y] &10$^{4.33}$ &10$^{1.97}$ \\
$10^5$     &  0.50  & -  &  - & yes & iPtot & Model A & 3e-2  & 39.2[y] &10$^{4.24}$ &10$^{1.84}$ \\
$10^5$     &  0.50  & -  &  - & yes & iPtot & Model A & 5e-2 & 42.6[y] &10$^{4.04}$ &10$^{1.81}$ \\
$10^5$     &  0.50  & -  &  - & yes & iPtot & Model A & 7e-2 & 42.2[y] &10$^{3.66}$ &10$^{1.44}$ \\
$10^5$     &  0.50  & -  &  - & yes & iPtot & Model A & 9e-2 & 21.8[y] &10$^{2.91}$ &10$^{0.72}$ \\
$10^5$     &  0.50  & -  &  - & yes & iPtot & Model A & 1e-1 & 16.88[y] &10$^{2.8}$ &10$^{0.53}$ \\
$10^5$     &  0.50  & -  &  - & yes& iPtot & Model A & 1.2e-1 & 10.86[y] &10$^{2.64}$ &10$^{0.36}$ \\
$10^5$     &  0.50  & -  &  - & yes& iPtot & Model A & 1.5e-1 & 5.67[y] &10$^{2.33}$ &stable \\
$10^5$     &  0.50  & -  &  - & yes& iPtot & Model A & 1.7e-1 & 4.65[y] &10$^{2.11}$ &stable \\
$10^5$     &  0.50  & -  &  - & yes & iPtot & Model A & 2e-1 &stable&-&-\\
$10^5$    &  0.50  & -  &  - & - & iPtot & Model B &  1e-2 &27.27[y] &10$^{4.41}$ &-\\
$10^5$   &  0.50  & -  &  - & - & iPtot & Model B &  5e-2  &27.07[y] &10$^{4.41}$ &-\\
$10^5$   &  0.50  & -  &  - & - & iPtot & Model B &  1e-1  &29.70[y] &10$^{4.36}$ &-\\
$10^5$   &  0.50  & -  &  - & - & iPtot & Model B &  2e-1  & 32.31[y] &10$^{4.38}$ &-\\
$10^5$   &  0.50  & -  &  - & - & iPtot & Model B &  2.5e-1  &  33.89[y] &10$^{4.38}$ &-\\
$10^5$   &  0.50  & -  &  - & - & iPtot & Model B &  4e-1  &  37.56[y] &10$^{4.3}$ &-\\
$10^5$   &  0.50  & -  &  - & - & iPtot & Model B &  5e-1 &41.79[y] &10$^{4.25}$ &-\\
$10^5$    &  0.50  & -  &  - & - & iPtot & Model B &  7e-1 &48.84[y] &10$^{4.24}$ &-\\
$10^5$   &  0.50  & -  &  - & - & iPtot & Model B &  8e-1  & 41.31[y] &10$^{4.04}$ &-\\
$10^5$    &  0.50  & -  &  - & - & iPtot & Model B &  9e-1 &24.66[y] &10$^{3.22}$ &-\\
$10^5$   &  0.50  & -  &  - & - & iPtot & Model B &  1.0 &9.2[y] &10$^{2.94}$ &-\\
$10^5$   &  0.50  & -  &  - & - & iPtot & Model B &  1.1 &stable&-&-\\
$10^5$    &  0.50  & -  &  - & yes & iPtot & Model B &  1e-2 &34.04[y] &10$^{4.34}$ &10$^{2.21}$\\
$10^5$    &  0.50  & -  &  - & yes & iPtot & Model B &  5e-2 &35.44[y] &10$^{4.33}$ &10$^{1.91}$\\
$10^5$   &  0.50  & -  &  - & yes & iPtot & Model B &  1e-1  &36.56[y] &10$^{4.35}$ &10$^{1.96}$\\
$10^5$   &  0.50  & -  &  - & yes & iPtot & Model B &  2e-1  &41.31[y] &10$^{4.3}$ &10$^{1.82}$\\
$10^5$    &  0.50  & -  &  - & yes & iPtot & Model B &  2.5e-1 &  42.26[y] &10$^{4.29}$ &10$^{1.84}$\\
$10^5$    &  0.50  & -  &  - & yes & iPtot & Model B &  4e-1 &    47.56[y] &10$^{4.28}$ &10$^{1.8}$\\
$10^5$   &  0.50  & -  &  - & yes & iPtot & Model B &  5e-1 &51.16[y] &10$^{4.24}$ &10$^{1.83}$\\ 
$10^5$    &  0.50  & -  &  - & yes & iPtot & Model B &  7e-1 &59.42[y] &10$^{4.04}$ &10$^{1.68}$\\ 
$10^5$    &  0.50  & -  &  - & yes & iPtot & Model B &  8e-1 &48.16[y] &10$^{3.39}$ &10$^{1.13}$\\ 
$10^5$    &  0.50  & -  &  - & yes & iPtot & Model B &  9e-1 &24.45[y] &10$^{3.04}$ &10$^{0.56}$\\ 
$10^5$   &  0.50  & -  &  - & yes & iPtot & Model B &  1.0 &18.25[y] &10$^{2.72}$ &stable\\
$10^5$   &  0.50  & -  &  - & yes & iPtot & Model B &  1.1 &stable&-&-\\ \hline \hline

\end{longtable}
\footnotesize{{\sc Notes.} Columns are as follows: (1) Black hole mass (2) accretion rate (3)  the inner ADAF location (4) the outer radius location (5) presence of corona (6) viscosity type (7) magnetic field model used in the simulation (see Sec. \ref{sec:modela_modelb}), (8)  magnetic field coefficient used in simulation, (9) period of the outbursts, (10) amplitude of the disk outburst, and (11) amplitude of the corona outburst.}
\end{landscape}
\clearpage
\twocolumn

\clearpage
\onecolumn
\begin{landscape}
\begin{longtable}{c|c|c|c|c|c|c|c|c|c|c}
\caption{The list of models calculated in the present paper for 10$^7$ $M_{\odot}$.} \label{tab:modele-10-7-xicor-0}
\\
$M_{BH}$ & $\dot m$ & \textbf{ $R_{ADAF}$} & \textbf{$R_{out}$} & corona & viscosity & magnetic & magnetic & period& amplitude&amplitude\\ 
$[M_{\odot}]$ & [$\dot M_{Edd}$] & [$R_{Schw}$] & [$R_{Schw}$] &  & type &  & coefficient & & disk&corona\\ \hline

\endhead

$10^7$    &  0.20  & -  &  100 & - & iPtot &  -&- &50.6[y] &10$^{4.83}$ &-\\
$10^7$    &  0.20  & -  &  100 & yes & iPtot & - &- &52.6[y] &10$^{4.87}$ & 10$^{2.75}$\\
$10^7$    &  0.20  & -  &  50 & - & iPtot &  -&- &10.1[y] &10$^{4.79}$ &-\\
$10^7$    &  0.20  & -  &  50 & yes & iPtot & - &- &12.39[y] &10$^{4.97}$ &10$^{2.27}$\\


$10^7$    &  0.20  & -  &  100 & - & iPtot &  Model B&0.01 &52.76[y] &10$^{4.83}$ &- \\
$10^7$    &  0.20  & -  &  100 & - & iPtot &  Model B&0.05 &58.47[y] &10$^{4.78}$ & -\\
$10^7$    &  0.20  & -  &  100 & - & iPtot &  Model B&0.1 &69.7[y] &10$^{4.77}$ & -\\
$10^7$    &  0.20  & -  &  100 & - & iPtot &  Model B&0.2 &47.7[y] &10$^{4.77}$ &-\\ 
$10^7$    &  0.20  & -  &  100 & - & iPtot &  Model B&0.4 & 86.79[y] &10$^{4.77}$ & -\\
$10^7$    &  0.20  & -  &  100 & - & iPtot &  Model B&0.5 &106.06[y] &10$^{4.76}$ &- \\
$10^7$    &  0.20  & -  &  100 & - & iPtot &  Model B&0.7&151.08[y] &10$^{4.61}$ & -\\
$10^7$    &  0.20  & -  &  100 & - & iPtot &  Model B&0.8&174.67[y] &10$^{4.51}$ & -\\
$10^7$    &  0.20  & -  &  100 & - & iPtot &  Model B&0.9 &212.12[y] &10$^{4.36}$ & -\\
$10^7$    &  0.20  & -  &  100 & - & iPtot &  Model B&1.0 &235.04[y] &10$^{4.07}$ &-\\
$10^7$    &  0.20  & -  &  100 & - & iPtot &  Model B&1.1 &58.39[y] &10$^{2.81}$ &-\\
$10^7$    &  0.20  & -  &  100 & - & iPtot &  Model B&1.15 &22.35[y] &10$^{2.31}$ &-\\
$10^7$    &  0.20  & -  &  100 & - & iPtot &  Model B&1.25 &5.35[y] &10$^{0.56}$ &-\\
$10^7$    &  0.20  & -  &  100 & - & iPtot &  Model B&1.5 &stable& -&-\\

$10^7$    &  0.20  & -  &  100 & yes & iPtot &  Model B&0.01 &75.33[y] &10$^{4.66}$ &10$^{2.82}$ \\
$10^7$    &  0.20  & -  &  100 & yes & iPtot &  Model B&0.05&92[y] &10$^{4.83}$ &10$^{2.81}$ \\
$10^7$    &  0.20  & -  &  100 & yes & iPtot &  Model B&0.1 &107.58[y] &10$^{4.76}$ &10$^{2.65}$ \\
$10^7$    &  0.20  & -  &  100 & yes & iPtot &  Model B&0.2 &56.13[y] &10$^{4.39}$ &10$^{2.88}$ \\
$10^7$    &  0.20  & -  &  100 & yes & iPtot &  Model B&0.4 &90.46[y] &10$^{4.71}$ &10$^{2.83}$ \\
$10^7$    &  0.20  & -  &  100 & yes & iPtot &  Model B&0.5 &117.75[y] &10$^{4.79}$ &10$^{2.53}$ \\
$10^7$    &  0.20  & -  &  100 & yes & iPtot &  Model B&0.7
&127.75[y] &10$^{4.84}$ &10$^{2.01}$ \\
$10^7$    &  0.20  & -  &  100 & yes & iPtot &  Model B&0.8
&195.24[y] &10$^{4.54}$ &10$^{1.93}$ \\
$10^7$    &  0.20  & -  &  100 & yes & iPtot &  Model B&0.9 &228.57[y] &10$^{4.37}$ &10$^{1.83}$ \\
$10^7$    &  0.20  & -  &  100 & yes & iPtot &  Model B&1.0 &235.04[y] &10$^{4.07}$ &10$^{1.59}$ \\
$10^7$    &  0.20  & -  &  100 & yes & iPtot &  Model B&1.1 &65.59[y] &10$^{2.76}$ &10$^{0.48}$ \\
$10^7$    &  0.20  & -  &  100 & yes & iPtot &  Model B&1.15 &32.04[y] &10$^{2.33}$ &10$^{0.18}$ \\
$10^7$    &  0.20  & -  &  100 & yes & iPtot &  Model B&1.2 &17.53[y] &10$^{1.59}$ &-\\
$10^7$    &  0.20  & -  &  100 & yes & iPtot &  Model B&1.25 &stable&  -&-\\
$10^7$    &  0.20  & -  &  100 & yes & iPtot &  Model B&1.5 &stable& -&-\\ \hline \hline
\end{longtable}
\footnotesize{{\sc Notes.} Columns are as follows: (1) Black hole mass (2) accretion rate (3)  the inner ADAF location (4) the outer radius location (5) presence of corona (6) viscosity type (7) magnetic field model used in the simulation (see Sec. \ref{sec:modela_modelb}), (8)  magnetic field coefficient used in simulation, (9) period of the outbursts, (10) amplitude of the disk outburst, and (11) amplitude of the corona outburst.}
\end{landscape}
\clearpage
\twocolumn

\begin{figure*}[ht]
    \centering
        \includegraphics[scale=0.35]{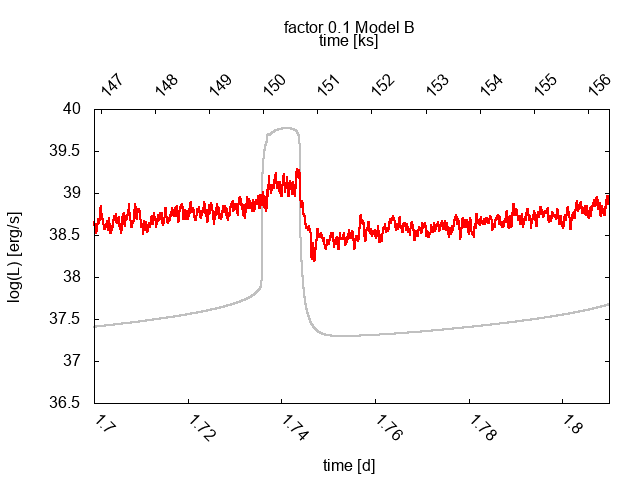}
             \includegraphics[scale=0.35]{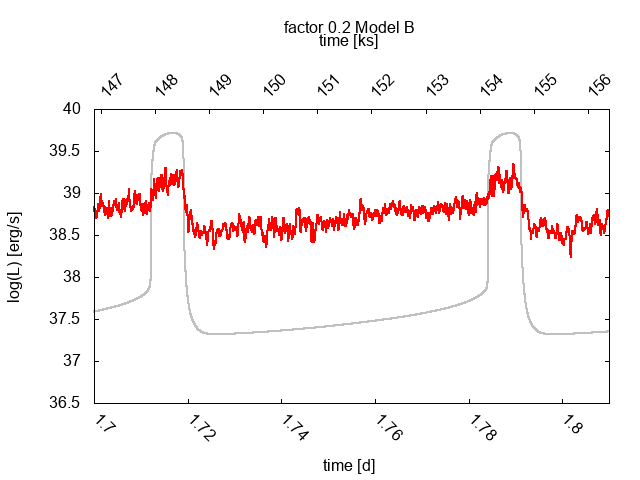}
      \includegraphics[scale=0.35]{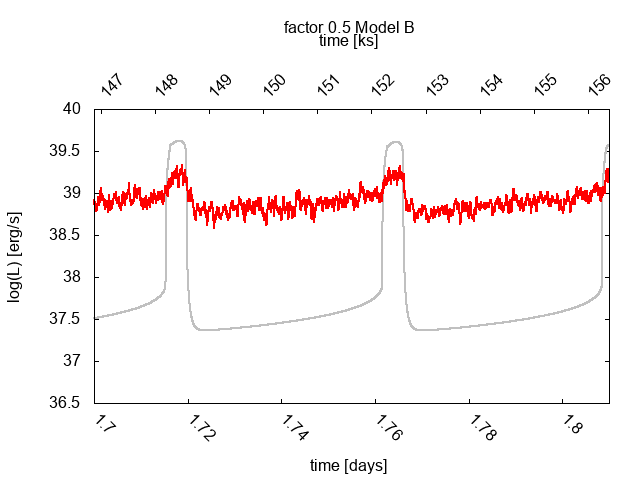}
      \includegraphics[scale=0.35]{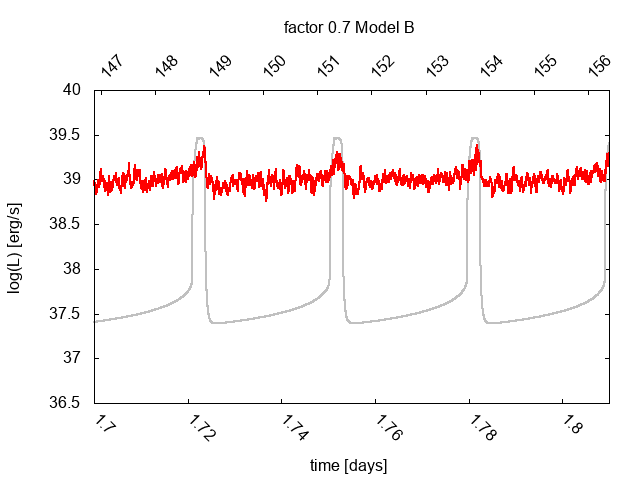}
            \includegraphics[scale=0.35]{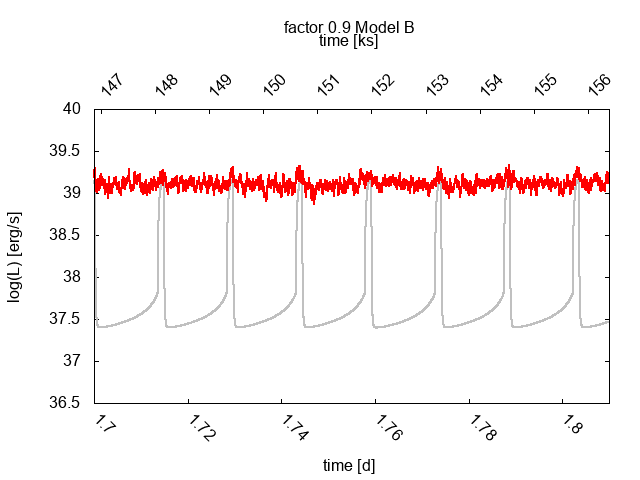}
             \includegraphics[scale=0.35]{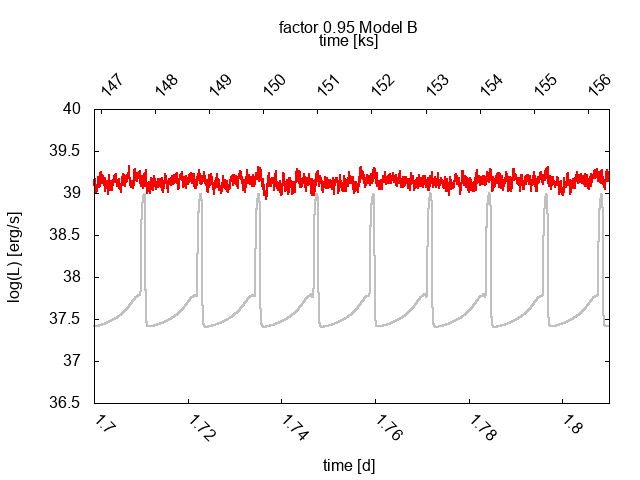}
               \includegraphics[scale=0.35]{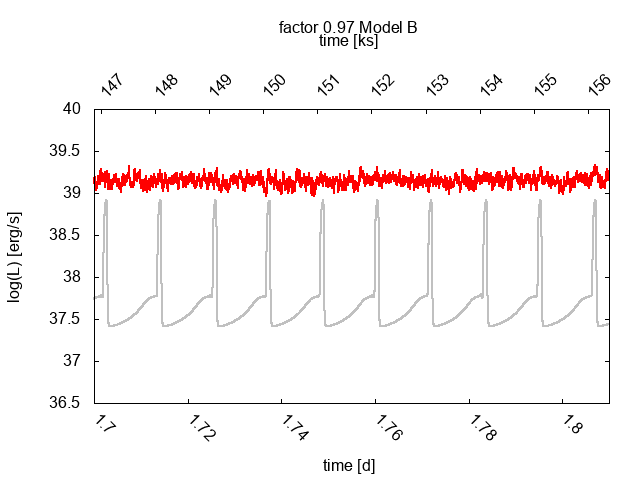}
               \includegraphics[scale=0.35]{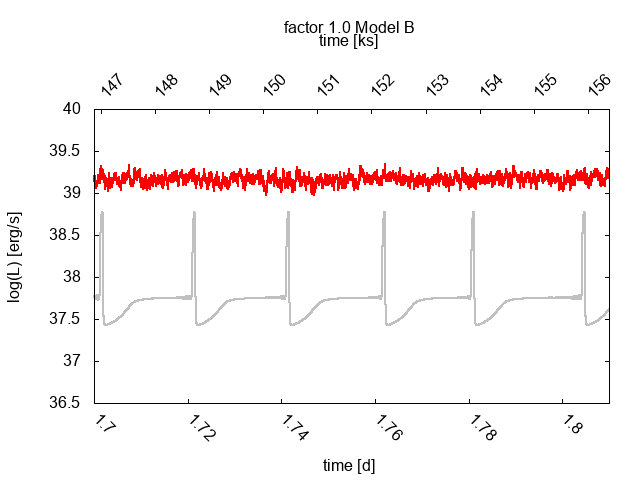}
       \caption{The disk (marked in gray) and corona (marked in red) lightcurves for 10M$\odot$
       for different coefficient value $b'$ for Model B (from the top left): $b'$ = 0 (base model),    0.1,  0.2, 0.5, 0.7, 0.9, 0.95, 0.97 and 1.0. For all cases the inner radius is 3 R$_{schw}$ and the outer radius R$_{out}$ = 300  R$_{schw}$.}
       
    \label{fig:mqsmodel2-iptot-corona-1-przez-czlon-xicor0}
\end{figure*}

\begin{figure*}[ht]
    \centering
 
           \includegraphics[scale=0.5]{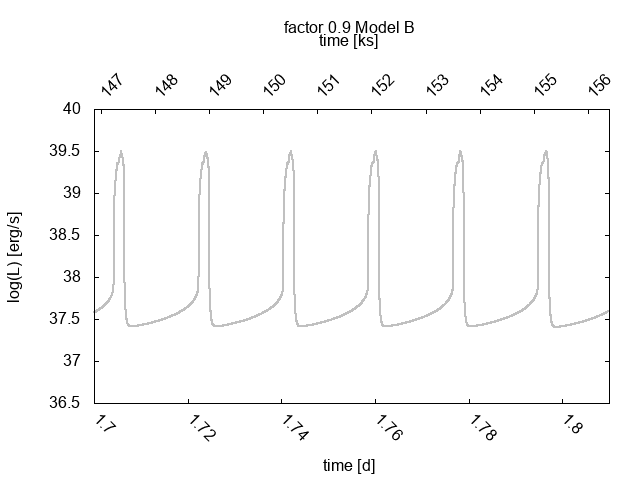}
  \includegraphics[scale=0.5]{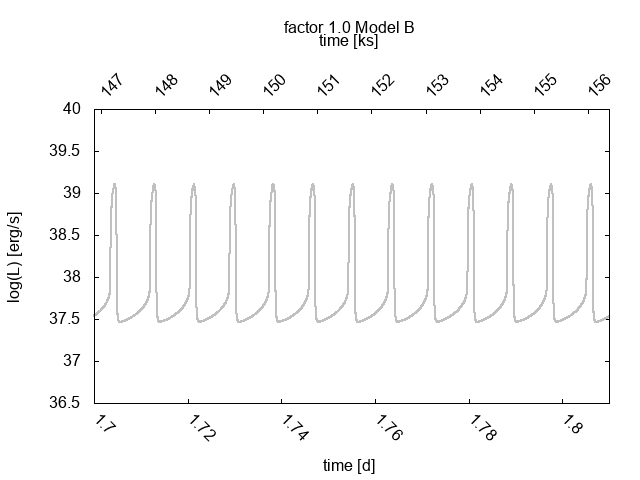}
       \caption{The disk (marked in gray) lightcurves for 10M$\odot$
       for different coefficient value $b'$ for Model B (from the top left):  0.9 (left) and 1.0 (right). For all cases the inner radius is 3 R$_{schw}$ and the outer radius R$_{out}$ = 300  R$_{schw}$.}
    \label{fig:mqsmodel2-iptot-disk-1-przez-czlon-xicor0}
\end{figure*}

Next, we use the same basic setup but assume Model B of the magnetic field effect on the disk structure (see Equation~\ref{eq:mag_flux_B}). We calculate again a sequence of models with increasing values of the parameter $b'$. We plot the models with the corona (see Figure~\ref{fig:mqsmodel2-iptot-corona-1-przez-czlon-xicor0}).  The initial trend with the rise of the magnetic field is very similar to the family A of models: the outbursts become shorter with the rise of the magnetic field term, the amplitudes become smaller, and the coronal outbursts have a smaller amplitude than the disk outbursts. For $b' > 0.7 $ outburst in the corona disappear, again replaced with purely stochastic variations. However, there is an important difference with Model A for the largest values of the parameter $b'$. When $b'$ is changed from 0.97 to 1.0, the amplitude decreases but the outburst duration increases from 16 minutes to 40 minutes, and the outburst shape changes, with an extended plateau developing before the consecutive outburst (see Figure \ref{fig:mqsmodel2-iptot-disk-1-przez-czlon-xicor0}).  The models without the corona follow the same pattern for small values of $b'$ (see outburst timescales and amplitudes in Table~\ref{tab:modele}. However, the last model ($b' = 1.0$) in the absence of the corona follows the trend of shortening the outburst scale (the outburst duration of 6 minutes), and no changes in the outburst shape are observed. Further increase of the parameter $b'$ leads to stable solutions.

Therefore, the role of the magnetic field is critical from the point of view of disk stability, outburst amplitudes, and timescales. Two proposed descriptions (Model A and B) imply similar behaviour, although with some differences in detail, as discussed.

\begin{figure*}[ht]
    \centering
 \includegraphics[scale=0.35]{plots/plots-mqso-iptot-czlon-factor-corona-xicor0/mqso-test-iptot-model2-27-10-2021-novis-wf0-adaptive-step-comparison-rout300-xicor0.png}
        \includegraphics[scale=0.35]{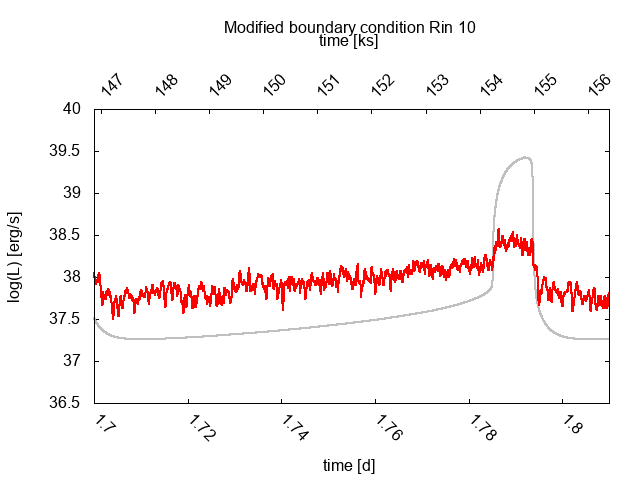}
         \includegraphics[scale=0.35]{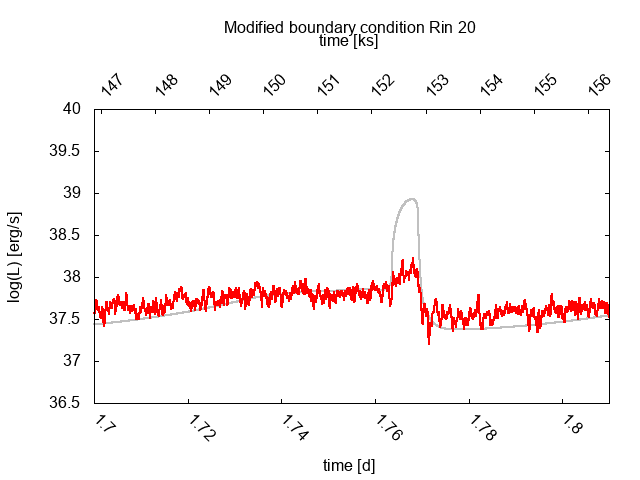}
       \caption{The disk (marked in gray) and corona (marked in red) lightcurve  for 10M$\odot$ with modified boundary condition (for details see Section \ref{sect:ADAF}) for different inner radii (R$_{ADAF}$) 3 (base model), 10 and 20 R$_{schw}$. The outer radius for all cases is R$_{out}$ = 300 R$_{schw}$.}

    \label{fig:ADAF3}
\end{figure*}

Next, we studied the dependence of the model on the presence of the inner ADAF hot flow. We used models of the disk with corona since in this case, the boundary condition requires that the coronal material flows as ADAF below $R_{ADAF}$, and models without corona cannot satisfy the boundary conditions as formulated in Section~\ref{sect:ADAF}. We do not include the magnetic field effect. The expectation was that perhaps the coronal flow and fast inner ADAF flow may combine towards shortening the outbursts. However, computations show that the outburst period slightly increases. The amplitude decreases as the unstable region shrinks, and for ADAF radius above 30, $R_{Schw}$ the instability ceases to exist. The trend is illustrated in Figure~\ref{fig:ADAF3}. Therefore, the inner ADAF can only damp the instability but does not affect the outburst timescale considerably.

In all previous computations, we assumed rather a high accretion rate, 0.67, appropriate for example for GRS 1915+105. We thus perform the test on how the models with somewhat lower accretion rates are modified. For $\dot m = 0.5$ and model B, without corona, we observe the same trend as for the higher accretion rate: as the parameter $b'$ rises, the outburst timescales shrink and the amplitudes go down. The trend is a little more shallow than for a higher accretion rate, so the unmodified model shows a shorter timescale at a lower accretion rate, but at $b' = 0.5$, both timescales are comparable.

In the case of a microquasar, we do not study the role of the outer radius because, in this case, the mass supply comes from the companion and the radius of the entire disk is large. In time-dependent computations, it is necessary to cover just the entire unstable zone, and no effects of the outer radius are expected.

\subsubsection{Intermediate Mass Black Holes}
\label{sect:IMBH}

\begin{figure*}[ht]
    \centering
          \includegraphics[scale=0.35]{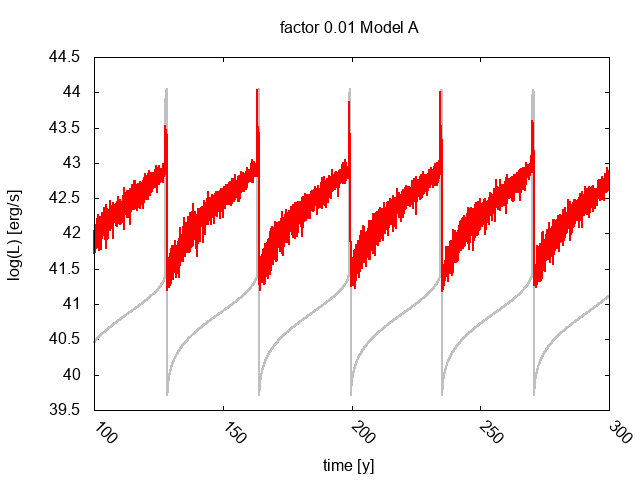}
          \includegraphics[scale=0.35]{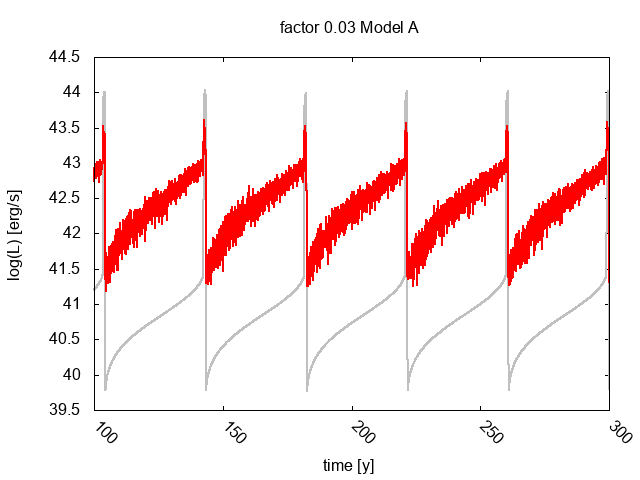}
          \includegraphics[scale=0.35]{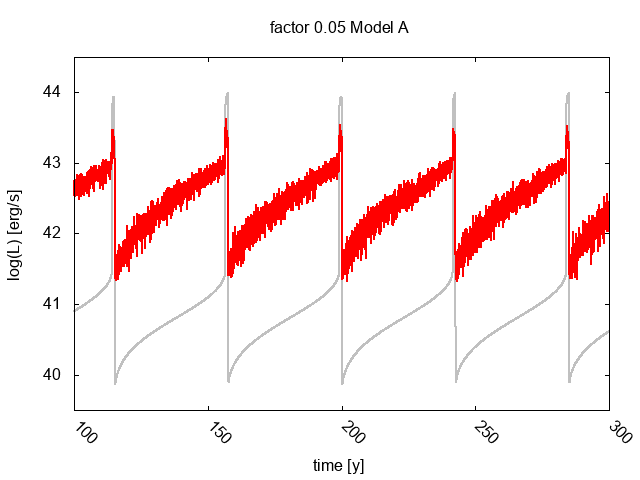}
          \includegraphics[scale=0.35]{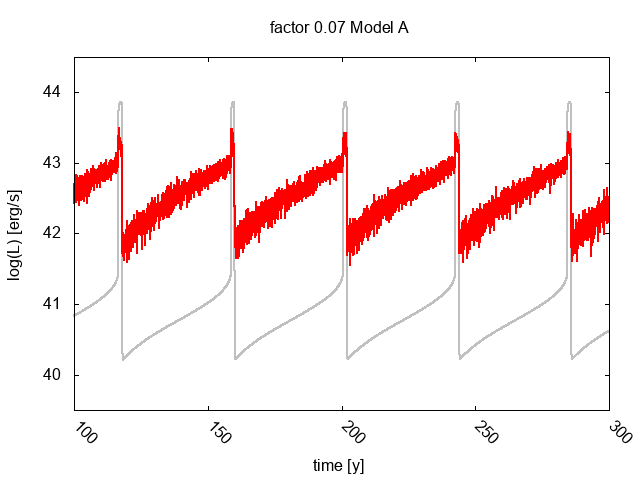}
           \includegraphics[scale=0.35]{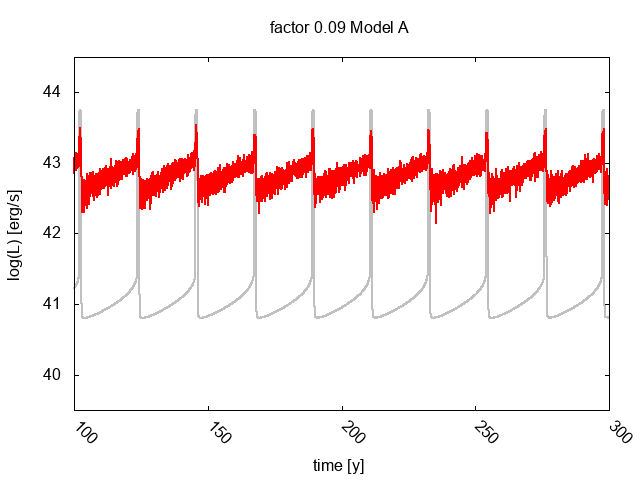}
             \includegraphics[scale=0.35]{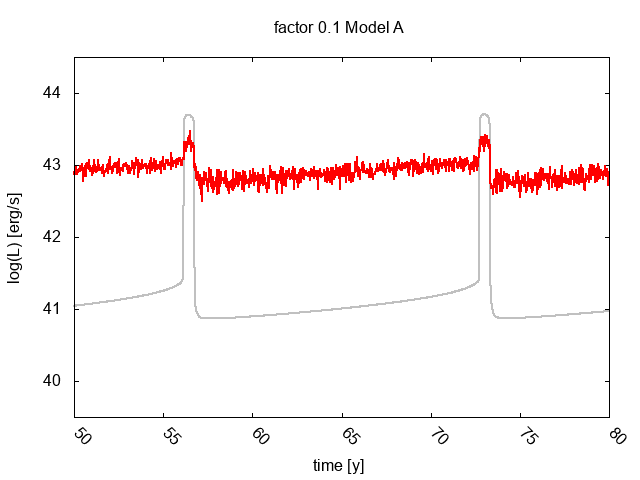}
              \includegraphics[scale=0.35]{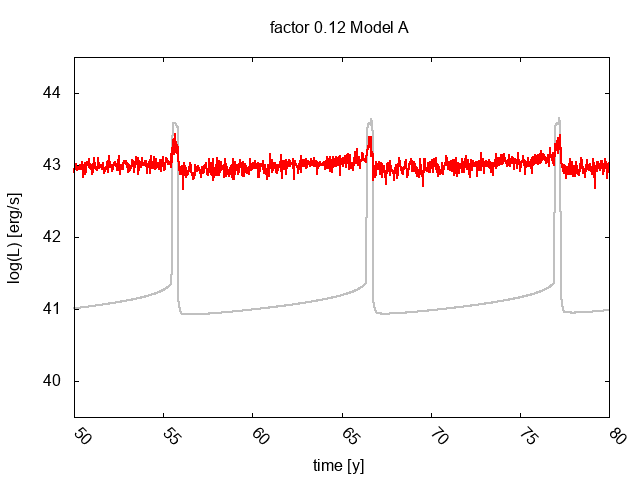}
               \includegraphics[scale=0.35]{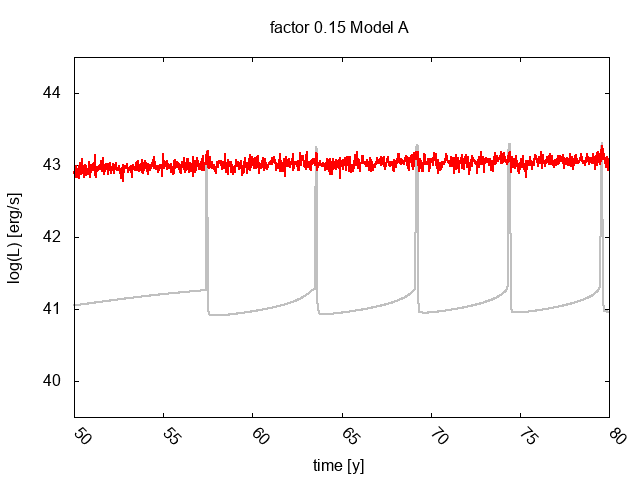}
                \includegraphics[scale=0.35]{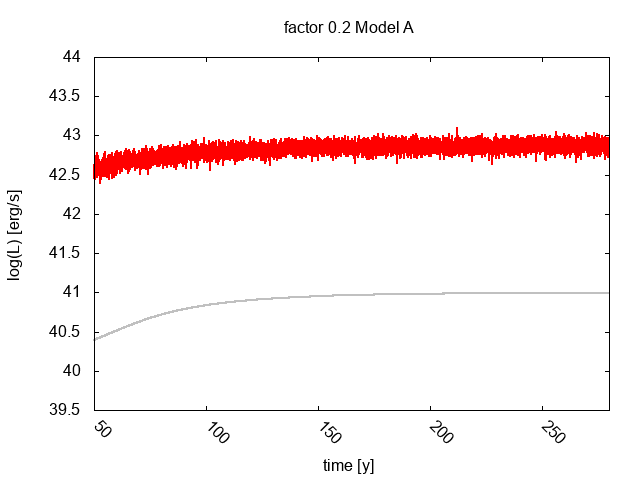}
 
        \caption{The disk (marked in gray) and corona (marked in red) lightcurve for 10$^5$M$\odot$  for different $b$ factors for Model A. The inner radius is 3 R$_{schw}$ and the outer radius is R$_{out}$ = 300 R$_{schw}$  for all cases presented in this plot.}
   \label{fig:gsn-modela-wm1}
\end{figure*}

\begin{figure*}[ht]
    \centering
          \includegraphics[scale=0.35]{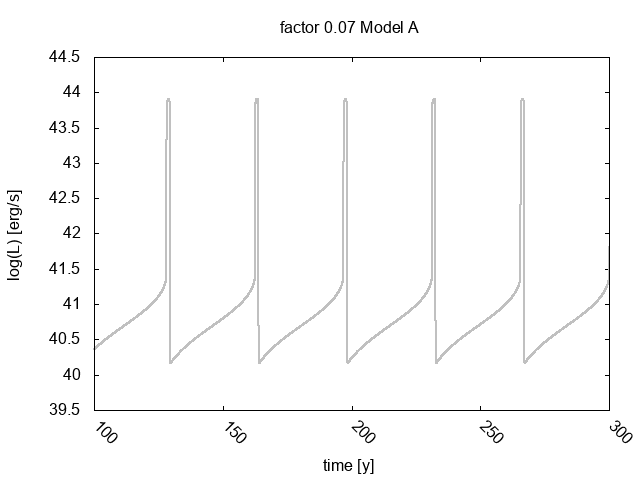}
          \includegraphics[scale=0.35]{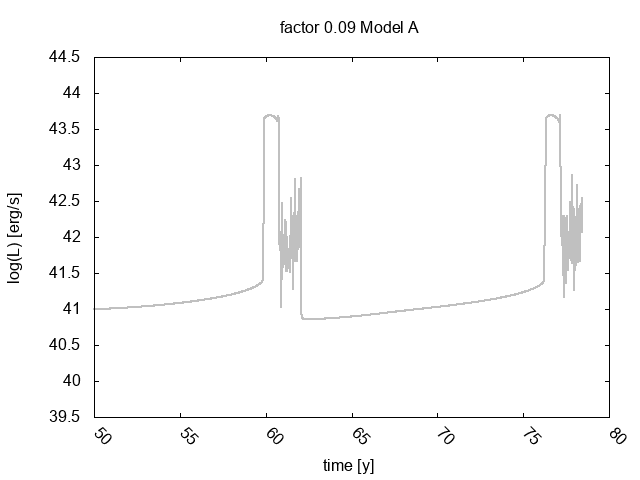}
      \includegraphics[scale=0.35]{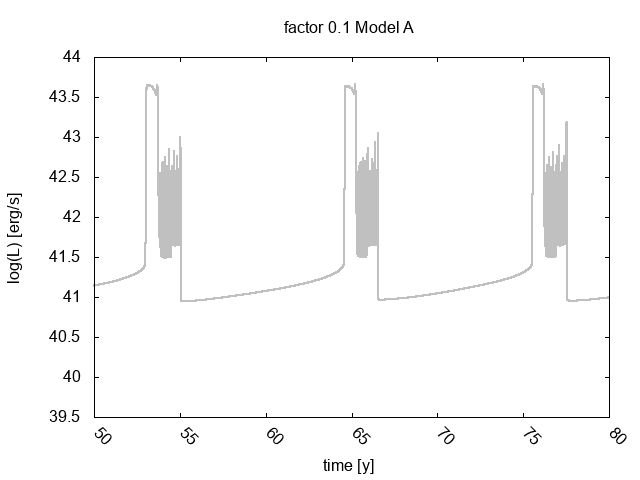}
         \includegraphics[scale=0.35]{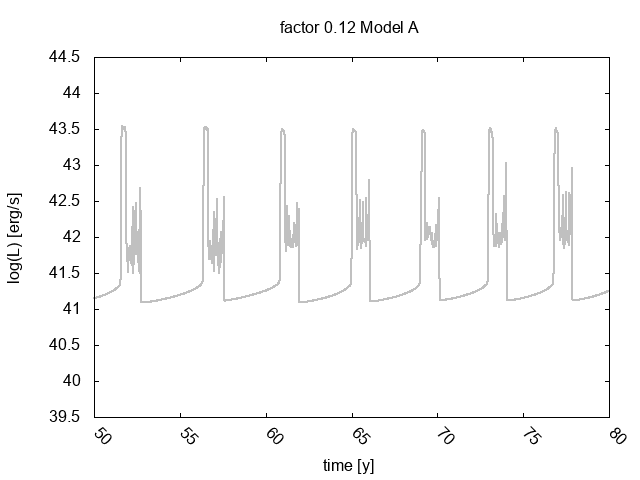}
         \includegraphics[scale=0.35]{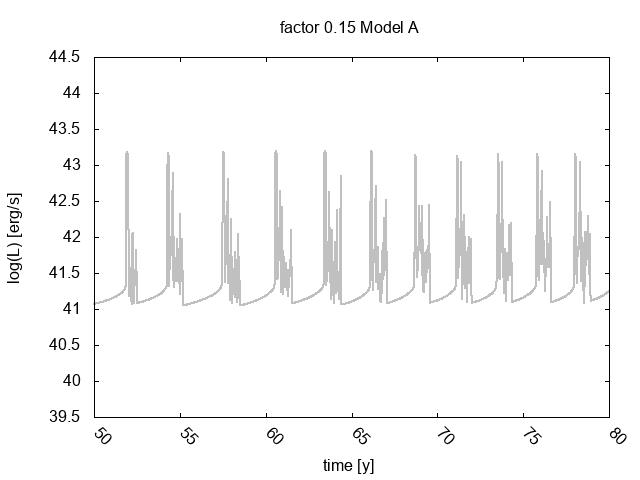}
       \includegraphics[scale=0.35]{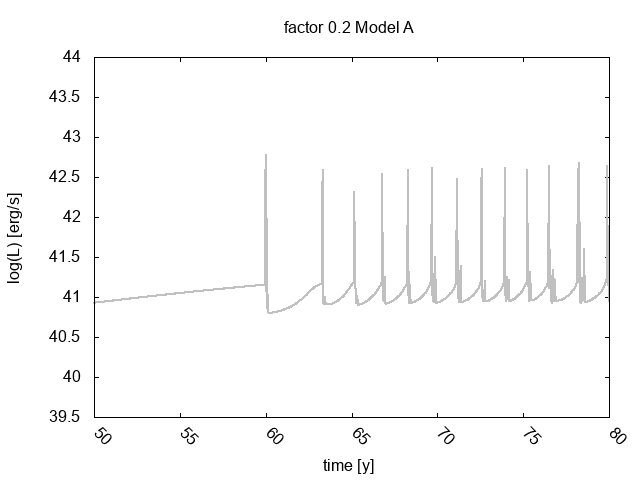}

    \caption{The disk (marked in gray) lightcurve for 10$^5$M$\odot$  for different $b$ factors for Model A. The inner radius  R$_{ADAF}$ = 3 and outer radius is R$_{out}$ = 300 R$_{schw}$  for all cases presented in this plot. Note different x-axis on plots.}
     \label{fig:gsn-modela}
\end{figure*}

In this section, we discuss the results for the black hole mass value $10^5 M_{odot}$ which is representative of sources like  HLX-1 or GSN  069. The model parameters are given in Table~\ref{tab:modele-10-5-xicor-0}. We usually set the value of the outer radius at $300 R_{Schw}$. We show the dependence of the disk and corona evolution on the importance of the magnetic field assuming model A in Figure~\ref{fig:gsn-modela-wm1}. We see that the characteristic timescales are much longer, in our basic model $b=$ the outbursts are 35.7 years, and in the absence of the corona, the duration of the limit cycle is only slightly shorter (27.6 yr). However, we observe again a strong trend of shortening the period with the increase of the magnetic field role, and for $b = 0.17$  the period reduces to 4.7 yr. Further increase of $b$ stabilizes the disk, and the disk without corona still shows outbursts for $b = 0.22$, and then the period is even shorter (1.1 yr). The shapes of the outbursts are different from those for the black hole mass $10 M_{\odot}$. The duration of the bright phase is much shorter in comparison to the overall duration of the limit cycle. They also show a dip and short-lasting minimum before the next outburst. However, a very interesting pattern appears if the accretion corona is not included. For $b=0.07$ outbursts of the disk without corona are still similar to outburst with corona, but for $b=0.09$, or higher, secondary oscillations develop right after the dominant peak (see Figure~\ref{fig:gsn-modela}). The interaction with the corona dumps this effect. However, similar complex substructure effects appear in the computations of larger masses quite frequently, and we will discuss that below. Overall, the sub-structure here does not affect the limit cycle duration.

In both cases, with and without a corona, we observe an interesting trend concerning the duration of the limit cycle. For very low values of the parameter $b$, the duration slightly rises with the increasing strength of the magnetic field, and only for $b \sim 0.07$ start to decrease very fast. No such effect was seen in the microquasar case. 

\begin{figure*}[ht]
    \centering
          \includegraphics[scale=0.35]{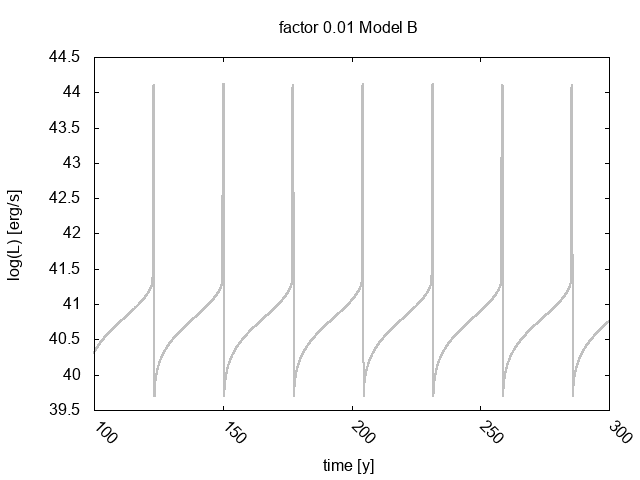}
    \includegraphics[scale=0.35]{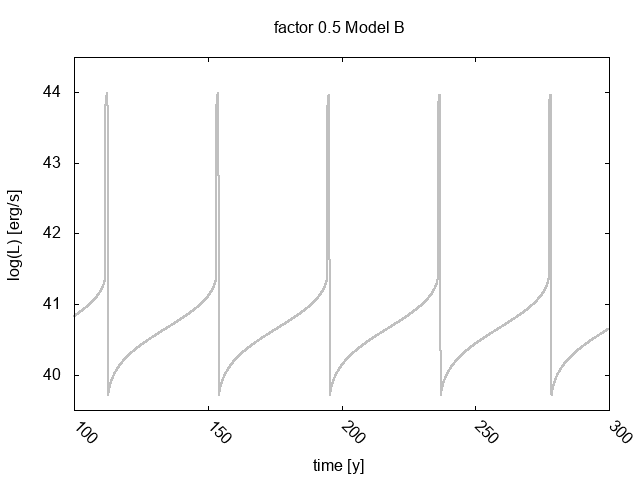}
      \includegraphics[scale=0.35]{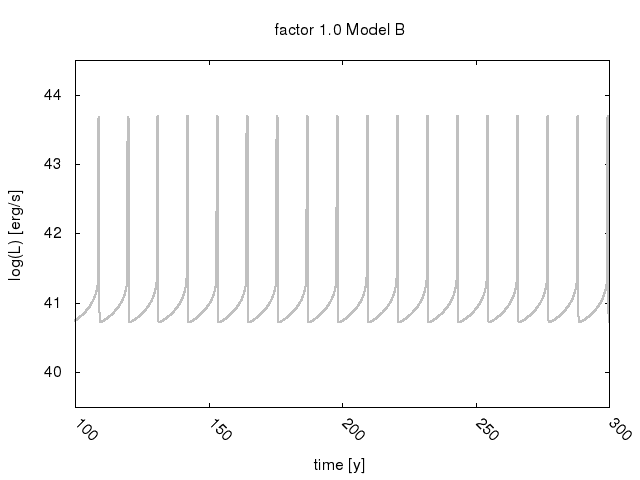}
   
    \caption{The disk (marked in gray) lightcurve for 10$^5$M$\odot$  for different $b'$ factors for Model B. The inner radius is 3 R$_{schw}$ and the outer radius is R$_{out}$ = 300 R$_{schw}$  for all cases presented in this plot.}
     \label{fig:gsn-modelb}
\end{figure*}

If we apply Model B to the description of the magnetic field, no substructure after the main peak develops. We see smooth single peaks, but with a similar profile as in the case of Model A - a very short bright phase, and a dip minimum after the peak (see Figure~\ref{fig:gsn-modelb}). Here we see an initial increase of the limit cycle duration with an increase in the magnetic field effect, later replace with a decrease, down to timescales of 9 years (without corona) and 18 years (with corona). The presence of the corona affects the disk behaviour quantitatively, but not qualitatively, in this case.


The effect of the inner ADAF also for this mass scale did not affect the solutions considerably. We calculated several models (see Table~\ref{tab:modele-10-5-xicor-0}), but as for the microquasar case, the timescale was weakly affected, only the amplitude dropped with the rising $R_{ADAF}$, and the model became stable for $R_{ADAF}$ above 30 $R_{Schw}$.

\begin{figure*}[ht]
    \centering
      \includegraphics[scale=0.5]{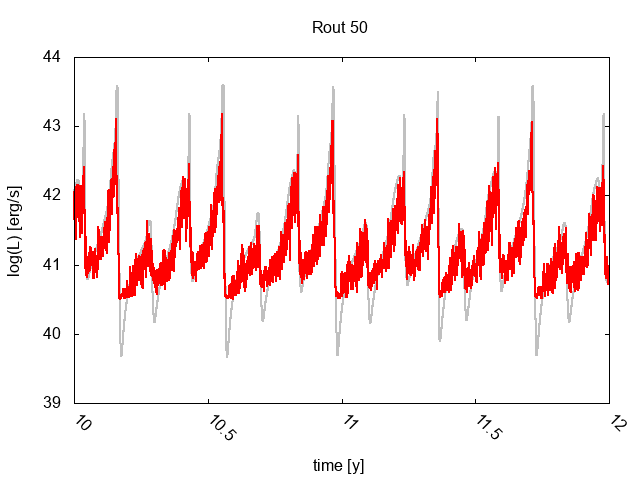}
            \includegraphics[scale=0.5]{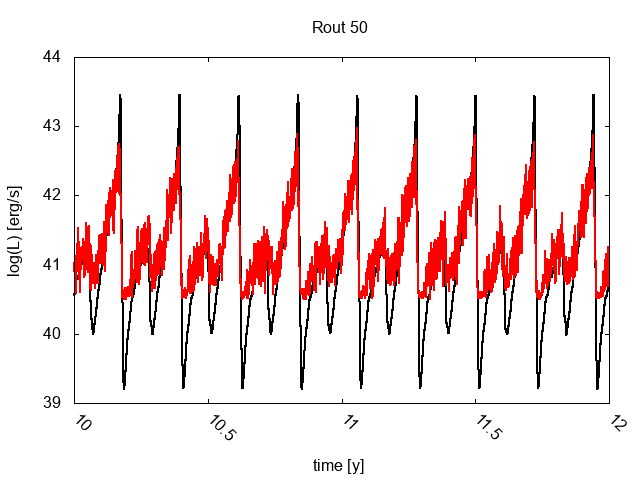}
            \newline
             \includegraphics[scale=0.5]{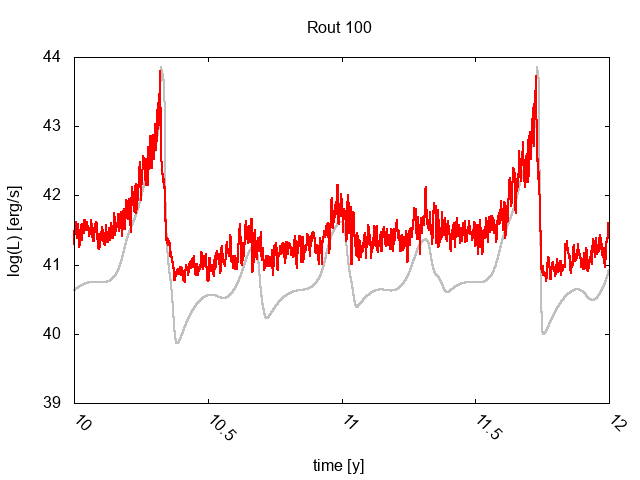}
          \includegraphics[scale=0.5]{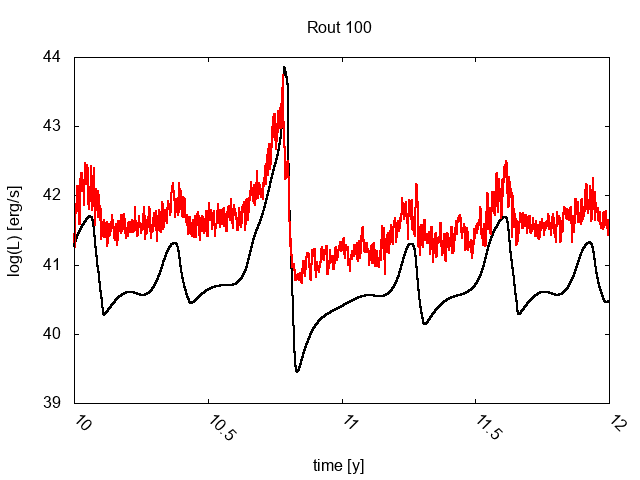}
   
    \caption{Upper panels: the lightcurves computed with the different number of grid points (left panel 98 points, right panel 34 points for the outer radius R$_{out}$ = 50 R$_{schw}$). Lower panels: 98  and 50 points for the outer radius R$_{out}$ = 100  R$_{schw}$. Model parameters in both cases: 10$^5M\odot$ and inner radius of 3 R$_{schw}$.}
     \label{fig:gsn-rout-50_1}
\end{figure*}

However, for the intermediate black hole mass scale we decided to check the role of the disk outer radius. The mass supply mechanism in the case of the intermediate black hole masses is not clear, and the source of mass can determine the outer boundary conditions.
Therefore, apart from the standard value of $R_{out} = 300 R_{Schw}$ we calculated models with the outer radii of $100 R_{Schw}$ and $50 R_{Schw}$, without and with the corona. The change of the outer radius has a clear and strong effect on the disk outbursts. Since now part of the potentially unstable disk is simply removed (although mass is still supplied there), we do not have the part responsible for the longest local evolutionary timescales. The global effect goes as expected: the overall period of the outburst shortens significantly, from 27 years down to 0.9 years for the smallest disk, without corona, and the presence of the corona does not change the trend. Since there is almost no difference in the disk behaviour between the disk wi and without corona, we plot only the case with a corona in Figure~\ref{fig:gsn-rout-50_1}, left top panel. Apart from the shortening of the outburst, we see an important change in the outburst character. 

Outbursts of disk with the outer radius 300 $R_{Schw}$ are almost identical. In the case of smaller disks (i.e. 50 $R_{Schw}$), we see a sequence of three outbursts of rising amplitude instead.
This behaviour is not surprising, and it was seen for example in computations of the outbursts of the cataclysmic variables caused by the ionization instability (see e.g. \citealt{Hameury1998}, their Figure 8.). Since \citet{Hameury1998} argued that the exact shape of such complex outbursts depends on the computational grid, we also tested the grid assumption. In Figure~\ref{fig:gsn-rout-50_1}, top right panel we plot the results obtained with smaller number of grid points, but coinciding for their position with the grid used in models with $R_{out} = 300 R_{Schw}$. We see that the exact sequence of sub-peaks is modified (two-peak sequence instead of three-peak sequence), but the complexity remains. We plot also the results for $R_{out} = 100 R_{Schw}$ (Figure~\ref{fig:gsn-rout-50_1}, lower panel). In this last case, the effect of the grid resolution change is small. 

\subsubsection{Supermassive black holes}

We saw from Section~\ref{sect:micro} and \ref{sect:IMBH} that the magnetic field shortens the duration of the outburst in the radiation pressure instability model. However, this effect is likely not strong enough if we aim at the possibility to model CL AGN. The standard outburst timescales in AGN are thousands of years \citep{czerny2009,grzedzielski2017}. On the other hand, as we already noticed in the case of IMBH, the position of the outer radius can lead to much shorter outbursts. The small outer radius is predicted if the activity is actually caused by the TDE effect. In the current study, we thus concentrate on the scenario where TDE is the underlying phenomenon, but the active evolutionary phase is long enough that radiation pressure instability turns on and a few cycles can be performed by the disk. We thus assume $R_{out} = 100 R_{schw}$, and we study the properties of such a model, including modifications caused by the magnetic field. This radius is much smaller than the whole instability zone which extends up to a few hundred $R_{Schw}$ \citep{janiuk2011} but in time-dependent computations, the regions above a thousand $R_{Schw}$ are still affected. Therefore, this is certainly the most important modification in comparison to the set of standard models discussed by \citet{grzedzielski2017}. All the models are listed in Table~\ref{tab:modele-10-7-xicor-0}. We assume a smaller accretion rate in these models, $\dot m = 0.2$ since CL AGN are rather not too close to the Eddington rate.

The reference model shows the outburst period of 51 yr. In \citet{grzedzielski2017} models with the viscosity law parameter $\mu = 0.56$, i.e. slightly higher than the sqrt law implied outburst periods of the order of 1000 years, and for the assumption of the torque proportional to the total pressure as currently used outbursts could not be calculated by \citet{grzedzielski2017} due to too long outbursts, too steep instability rise and computational problems. The choice of much smaller outer radii in the current paper reduces so much the outbursts' timescales, amplitudes, and the steepness of the thermal rise of the luminosity that no numerical problems are met. 

\begin{figure*}[ht]
    \centering
      \includegraphics[scale=0.25]{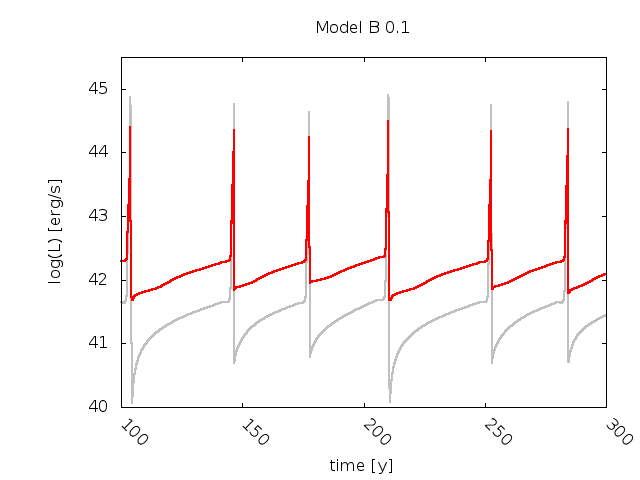}
         \includegraphics[scale=0.25]{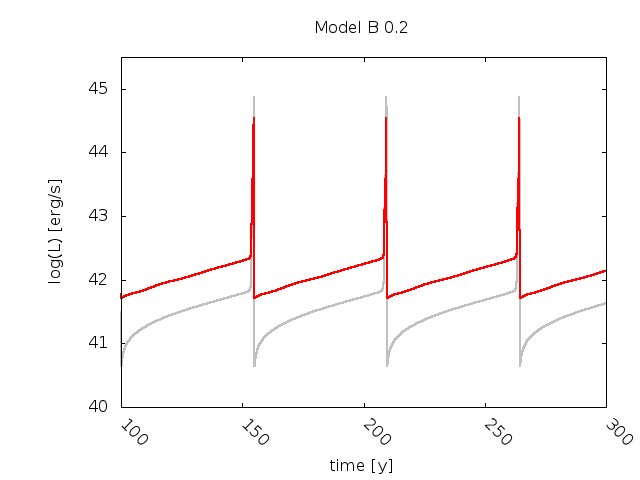}
    \includegraphics[scale=0.25]{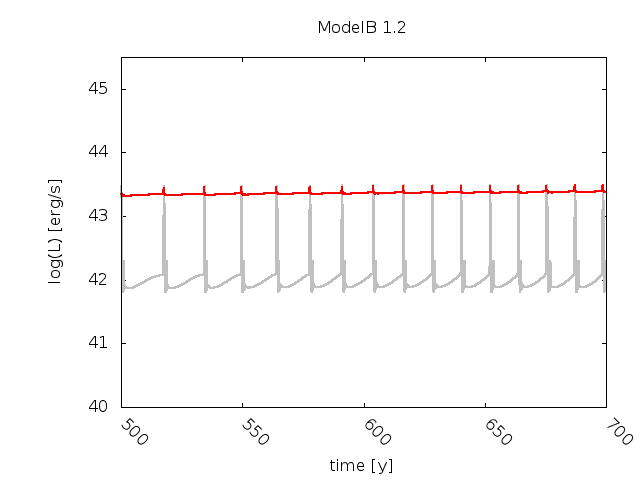}
        
        \caption{The disk (marked in gray) lightcurve for 10$^7$M$\odot$  for different $b'$ factors for Model B. The inner radius  R$_{ADAF}$ = 3 and outer radius is R$_{out}$ = 100 R$_{schw}$  for all cases presented in this plot.}
   \label{fig:agn-modela-dysl}
\end{figure*}

We analysed in detail further reduction of the outburst period which might be caused by the action of the magnetic field. We used Model B for this purpose. The results only weakly depended on the presence or absence of the corona. The trend was similar to what we noticed already for the IMBH case: the period was rising with the increasing strength of the magnetic field parametrized by the coefficient $b'$. However, for larger masses, this rising trend continued to much larger values of $b'$ and reversed to shortening of the limit cycle duration only for $b' > 1.0$. Then as before, for still larger values of $b'$ the period as well as the amplitude went down, and for $b' > 1.25$ the disk became stable. The trend is illustrated in Figure~\ref{fig:agn-modela-dysl}.

The model without a corona has similar properties to the model with the accreting corona. As for the IMBH case, the outbursts are very sharp, followed by a deep minimum. For example, the duration of the limit cycle for $b' = 1.0$ (without corona) lasts 235 years, but the duration of the bright phase is about 20 years. The influence of the grid resolution change on the results is noticeable, 
outbursts become shorter for smaller number of grid points. It is clearly visible for the case of disk with the outer radius  R$_{out}$ = 50 R$_{schw}$ (see Figure \ref{fig:gsn-rout-50_1}, upper panel),  in which period of outbursts decreased by a factor of 3.
Furthermore, the outbursts are not identical, but alternating between stronger and weaker outbursts.

\begin{figure*}[ht]
    \centering
    \includegraphics[scale=0.4]{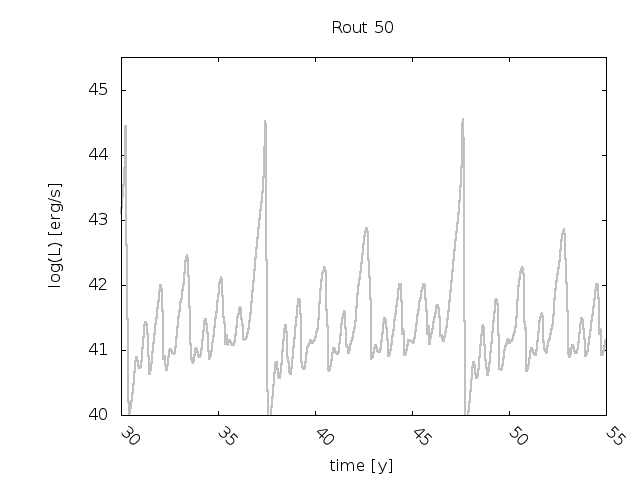}
      \includegraphics[scale=0.4]{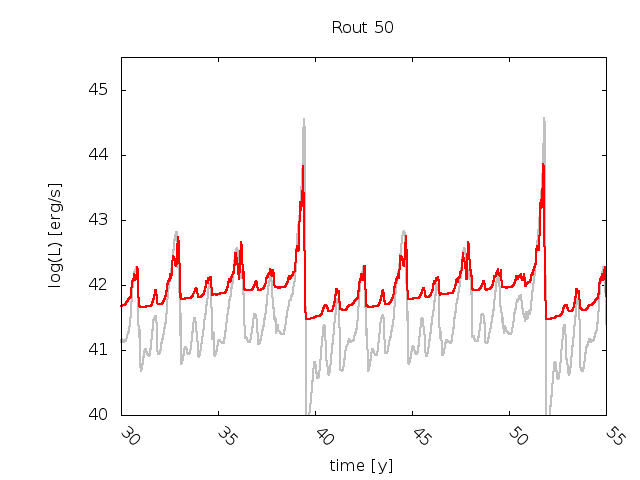}
        \caption{The disk (marked in gray) and corona (marked in red) lightcurves for 10$^7$M$\odot$  for the inner radius 3 R$_{schw}$ and outer radius is R$_{out}$ = 50 R$_{schw}$.}
        \label{fig:AGN_r50}
\end{figure*}

We checked the role of the inner ADAF in the case of massive black holes, but again it was not qualitatively affecting the results, apart from a systematic decrease of the outburst amplitude with the rising $R_{ADAF}$. However, the role of the outer radius is dominant, so we also tested the case of a still smaller value, $R_{out} = 50 R_{Schw}$. In this case, we observe a similar phenomenon as for the IMBH: the time evolution became very complicated, with a sequence of outbursts of different heights repeating regularly. The effect is only weakly affected by the presence or the absence of the corona (see Figure~\ref{fig:AGN_r50}).

\subsubsection{Global trends across the mass scale and the magnetic field strength}

In the three previous subsections, we discussed the dependencies of the outburst properties separately for each black hole mass. We showed that the period has a complex non-monotonic dependence on the strength of the magnetic field for higher masses. To show the pattern more clearly, we plot the global output parameters - just the amplitude and the duration - for all three black hole mass values next to each other, for model B. Since the effect of the corona on the disk outburst parameters is weak, we only plot models without corona.

\begin{figure*}
   \includegraphics[scale=0.5]{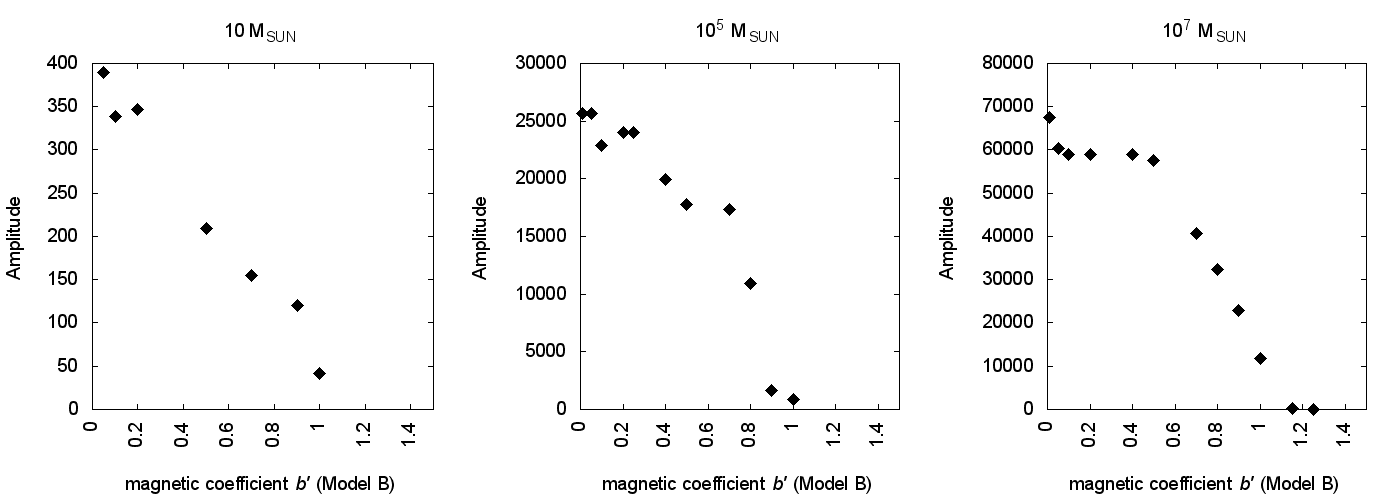}
   \includegraphics[scale=0.5]{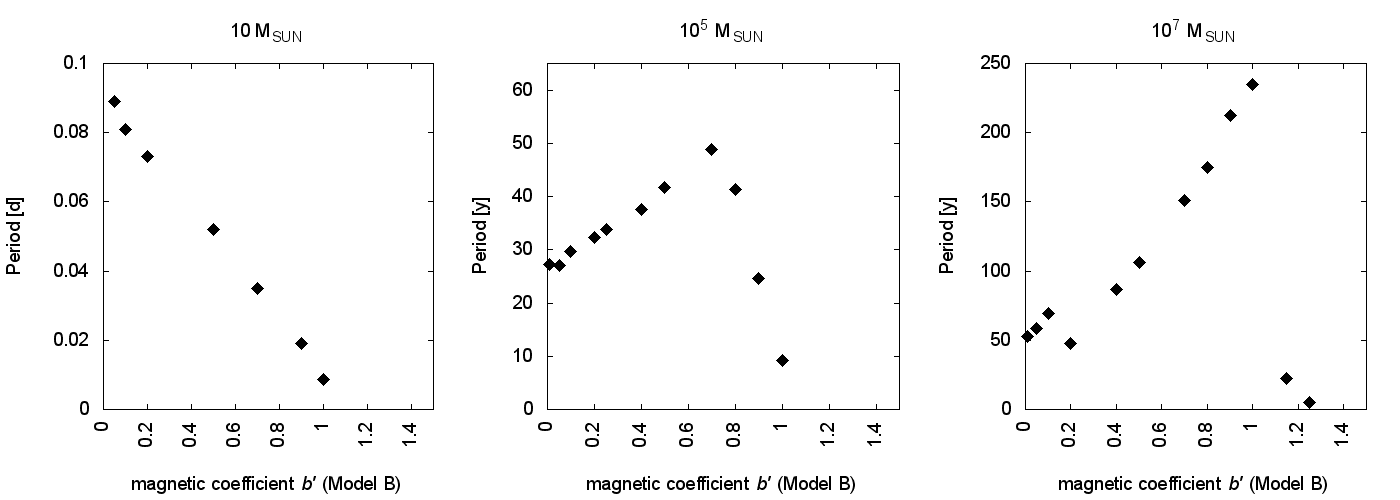}

             \caption{Dependence between the amplitude of the outburst (top panel) and the duration of the limit cycle (lower panel) of the accretion disk and magnetic coefficient $b'$ for the black hole masses $10 M_{\odot}$ (left panel), $10^5 M_{\odot}$  (middle panel), and $10^7 M_{\odot}$  (right panel).
             }
    \label{fig:modelb-all-masses-factors-both}
\end{figure*}

Figure~\ref{fig:modelb-all-masses-factors-both} shows that the amplitude of the outburst decreases with the field strength, independently of the black hole mass. However, in the case of the period we see a clear trend in the pattern. In the case of the IMBH, the period is rising first, and then falling. In microquasars we see only the second, decreasing branch while for a supermassive black hole the first, rising branch dominates, and the shortening of the period happens for very high values of $b'$, and then the change is very rapid. 

Looking at these plots we see that the amplitudes of the outbursts are rather large, and they rise with the black hole mass. In the case of microquasars, the observed amplitudes in GRS 1915+105 are of the order of 3 to 16 in the heartbeat states \citep[][]{belloni2000}. The observed timescales of these outbursts range from 40 s to 1500 s, and they weakly correlate with the amplitude. In our set of models, we have such solutions if the effect of the magnetic field is strong. For example, $b > 0.09$ in Model A without corona gives the right period range, and $b > 1.2$ gives also the right amplitude. Model B predicts amplitudes that are a little too high for the minimum period seen in the solutions. Another microquasar, IGR J17091–3624, has amplitudes up to the factor of 20, and the outburst timescales cover the range from 2 s to 100 s \citep{2011Altamirano}, on average shorter than in GRS 1915+105. It might be related to a somewhat lower value of the Eddington rate and a mass in this source. Our grid of models does not generally cover densely the mass and Eddington ratio grid but the solution for $\dot m = 0.5$ shows the period shorter than the model with $\dot m = 0.67$ (0.06 vs. 0.09 day) for the same value of $b'$. It is also important to note that a relatively small further increase in the strength of the magnetic field stabilizes the disk, and this may be consistent with the fact that heartbeat states are not always present. 
We also decided to check the importance of black hole mass and $\dot m$ assumed in our modelling. 
In the case of $10 M_{\odot}$ we see a monotonic decrease of the period, whereas for $10^5$ and $10^7 M_{\odot}$ we notice decreasing and increasing along the $b$ for the model A. Models differ from each other with two parameters: mass and accretion rate. To determine which of those is the leading parameter we computed a few cases for  $10 M_{\odot}$ with accretion rate $\dot m = 0.5$ (the same as used for $10^5 M_{\odot}$). For this simple test, we notice a similar monotonic decrease in the period. Thus, we claim that black hole mass is the leading parameter in this trend.

The comparison with the data is not precise since in the data we measure the amplitude as the count rate, and that depends strongly on the selected energy band while in models we predict the bolometric luminosity of the disk. Also, if the corona luminosity is added to the disk, the amplitudes measured from the model can be smaller since the accreting corona has much lower outburst amplitudes than the disk. 

In application of the model to IMBH, we can refer to observations of the outbursts in the object HLX-1 which shows regular outbursts with a timescale of 400 days, the amplitude of about a factor of 100 \citep{2015Yan},  and which was already successfully modeled as the effect of the radiation pressure instability \citep{2016Wu,grzedzielski2017}. Solutions with these properties are characterized by the magnetic field strength of $b \sim 0.2$ (Model A). The timescales in Model B are somewhat longer, and they rapidly become shorter for $b'$ between 1.0 and 1.1, when the model becomes stable. This leaves a very narrow range of potential parameters.

\begin{figure}
   \includegraphics[scale=0.4]{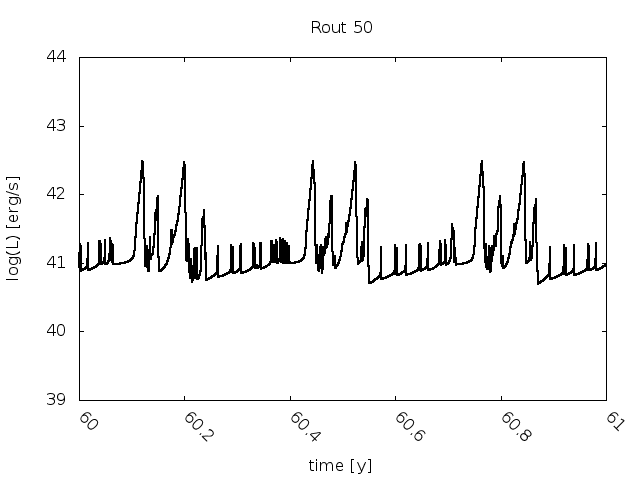}
             \caption{The disk (marked in black) lightcurve for 10$^5$M$\odot$  for  $b'$ = 0.22 factors for Model A. The inner radius 3 R$_{schw}$ and outer radius is R$_{out}$ = 50 R$_{schw}$.}
             
    \label{fig:mag_and_small_raout}
\end{figure}

However, a new class of outbursts now known as Quasi-Periodic Eruptions (QPE) has been detected and the black hole masses in these sources are in the range of IMBH or low mass AGN ($4 \times 10^5 M_{\odot}$ in GSN 069, \citealt{Miniutti2019}, $(0.8-2.8) \times 10^6 M_{\odot}$ in RX J1301.9+2747, \citealt{giustini2020}, low but highly uncertain masses in  eRO-QPE1 and eRO-QPE, \citealt{arcodia2021}). The timescale there is much shorter, of the order of hours. Our results presented in Table~\ref{tab:modele-10-5-xicor-0}, for the outer radius of $300 R_{Schw}$ show the minimum of the limit cycle timescale of 1 year (Model A, $b = 0.22$, no corona). Models with the smallest outer radius used in our computations ($50 R_{Schw}$) implied a shortening of the outburst time by a factor of 30, by a factor of 2 more than the reduction of the timescale expected from the simplest scaling of the dynamical time with the radius (power 3/2). Therefore, recalculating the model mentioned above with $b = 0.22$ might shorten the outburst timescale down to 10 days. We performed such calculations and the result is presented in Figure~\ref{fig:mag_and_small_raout}. The period of the short small outbursts is about 8 days, but the object's luminosity is systematically rising, and small outbursts are accompanied by large outbursts. So short timescales are possible to achieve in our model, and further decrease of the outer radius can give still shorter timescales. The duration of the bright phase is short in comparison with the duration of the limit cycle, as observed in the QPE phenomena. However, we cannot claim yet full success in explaining QPE since the overall pattern may not be as expected, and in the future, a much more careful approach to the outer boundary condition will be needed. 

However, our results show clearly that short timescale (hours) regular outbursts are consistent with the radiation pressure instability {\it only} in the case of very small outer disk radii. This means that the radiation pressure instability model can be applied to QPE events only under two conditions: (i) the general rise in the source activity must be due to the TDE effect, as this would limit the outer disk radius in a natural way  (ii) the role of the magnetic field in the disk must be large. In the current paper, we did not aim at finding unique parameters for QPE sources since the black hole mass measurements in these sources are quite uncertain, but the conditions formulated above are generic. The QPE sources detected so far were indeed activated by TDE as they happened in previously non-active galaxies.

In larger black hole masses characteristic of typical AGN we only considered TDE powered events since we aimed as testing whether rapid CL events can be explained by radiation pressure instability. The observed timescales in CL are not well constrained since multiple events are rarely detected, and in the case of a single transition event the process is usually not well captured. However, the typical timescales range from months to years, which may also be a selection bias against longer timescale changes that cannot be followed by the optical instruments. Timescales as short as that require again a combination of the small outer radius (its reduction from 100 to 50 $R_{Schw}$ reduces the period by a factor of 10) and the presence of the strong magnetic field (factor $b'$ larger than 1.2 by itself gives the periods of 5 years and shorter). The required parameter range is rather narrow, a slight rise of $b'$ stabilizes the disk. 

Such a small outer radius of the active accreting disk may pose a problem to the broad band SED modelling and the formation of the BLR which is present in CL AGN. Since the typical appearance of the BLR requires the presence of the irradiated material at a few hundred - few thousands of $R_{Schw}$ such a CL AGN must have a much longer history of typical AGN activity in the past. We address this issue in more detail in the Discussion.

\section{Discussion}

We use the modified version of the time-dependent code GLADIS (Global Accretion Disk Instability Simulation) developed originally by \citet{2002janiuk} to analyse the potential role of the radiation pressure instability in various objects across the broad range of the mass scale. The new modifications included the presence of the inner ADAF flow, the presence of the strong magnetic field studied by \citet{2015begelman}, and the option to constrain the disk size to the small outer radius in the case of IMBH and AGN. We used the disk plus hot corona flow model, previously studied by \citet{2007janiuk}.

The presence of the inner ADAF did not affect considerably the time evolution period of the disk. In \citet{2020A&A...641A.167S} the argument was raised that the narrow instability zone might lead to shorter timescales of the outburst but the numerical computations demonstrated that the instability zone is always broad when the limit cycle operates, and the effect of the inner ADAF is relatively unimportant unless it stabilizes the disk. 

The accreting corona also does not affect the disk evolution timescale considerably. We only note that in all solutions the corona follows the disk outburst but the amplitude of this outburst is smaller than the disk amplitude. Our model only predicts the bolometric luminosity (separately for the disk and for the corona), so we cannot address here directly the amplitudes measured in specific energy bands in the observational data. However, two other new parameters have a very strong effect on the time evolution of the disk.

The first of these parameters is the strength of the magnetic field. We include it following the previous analytical studies by \citet{2015begelman} and \citet{czerny2003}. We adopt ways of parametrizing the role of the magnetic field, described as Model A (see Equation~\ref{eq:mag_flux_A}) and Model B (see Equation~\ref{eq:mag_flux_B}). We expected that the energy transport mediated by the magnetic field will act towards stabilizing the disk by reducing the amplitude and shortening the limit cycle period), as implied by previous simple parametric studies by \citet{1994ApJ...436..599S}. However, our global numerical computations show that such a monotonic behaviour happens only for the microquasar case of black hole mass $10 M_{\odot}$. The strength of the magnetic field being an arbitrary parameter gives enough freedom to model outbursts seen in heartbeat states, the model set has a large range of predicted amplitudes and timescales. Of course, our results do not rule out previous successful attempts to model these states, based on modified viscosity law \citep{2000nayakshin} or wind/jet outflow \citep{2000janiuk,2000nayakshin,2015janiuk}.  

As was argued by \citet{2016Wu}, the radiation pressure instability operates at all mass scales, and after some adjustment, it can well explain the 400 days' outbursts in IMBH source HLX-1, or lasing hundreds to thousands of years of activity episodes in radio-loud AGN. In this last case, radiation pressure instability provides an explanation for an excess of short-lived sources \citep{czerny2009}. However, recently observed QPE events or repeating outbursts in CL AGN (like NGC 1566) represent orders of magnitude shorter timescales than provided by the {\it standard} radiation pressure instability models. Even the introduction of the magnetic field cannot change this conclusion.

Here comes the role of the last new parameter we introduced: the outer radius of the disk. In standard modeling \citep[see e.g.][]{2002janiuk,grzedzielski2017} one adopts the value of the outer radius $R_{out}$ large enough that the disk during the outburst remains stationary there, and in this case the specific value of the outer radius is irrelevant. Here, for IMBH and AGN, we considered disk outer radii which are much smaller. Since all the local timescales are rising with the disk radius \citep[e.g.][]{czerny2006}, a smaller disk outer radius shortens the timescales very efficiently and combining small $R_{out}$ and large values of magnetic field parameters $b$ or $b'$ we can shorten the outbursts down to hours/days for IMBH/AGN. However, introducing small $R_{out}$ has some direct and indirect consequences for the modeled scenario. 

Mathematically, in the current model, we assume that the accretion is at $R_{out}$ is constant, and the disk parameters there are fixed by the adopted external accretion rate. This radius forms an impenetrable barrier to any heating/cooling fronts which propagate inside the disk, and these cooling/heating waves reach the outer radius (since it is small, well within the instability zone) and get reflected. This increases the level of non-linearity in the equations, and most of the solutions with very small $R_{out}$ show complex multi-peak outbursts, characteristic of the first stage of the development of the deterministic chaos \citep[see e.g. discussion by][and the references therein]{grzedzielski_sukova2015,sukova2016}. However, the details of this phenomenon are certainly sensitive to the way how this outer boundary is set, and this in principle should be related to the global scenario which allows us to consider the small value of $R_{out}$ as representing the reality.

\begin{figure}
   \includegraphics[scale=0.3]{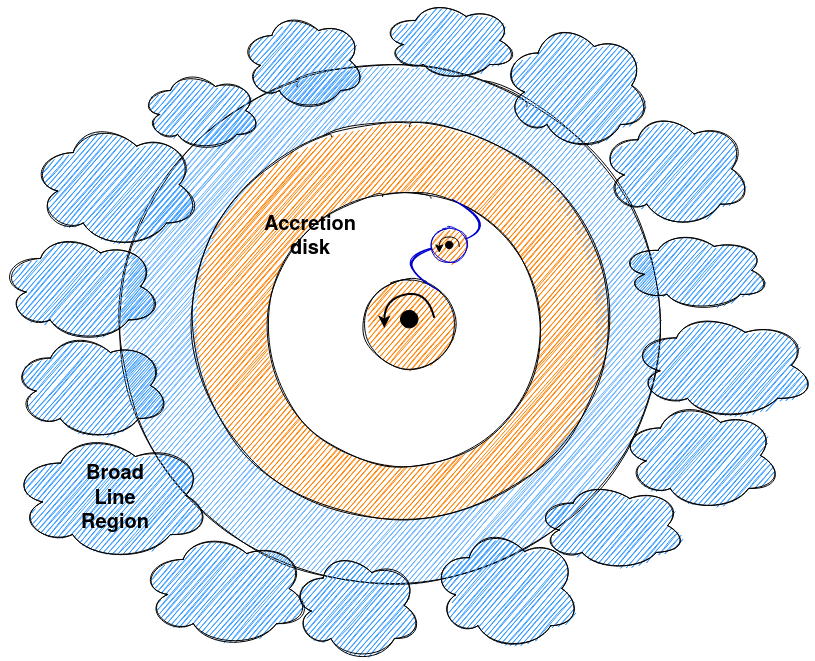}
             \caption{Schematic view of the scenario of black hole binary system with the smaller mass secondary black hole. The orange color represents the accretion disk (with a white gap caused by the presence of the secondary black hole and a smaller disk around it). The blue curve in the gap represents the material flowing through the gap. Blue clouds on the outer blue ring represent the broad line region, which is the source of flowing material.}
             
    \label{fig:tde_bbh}
\end{figure}

The first scenario is a compact TDE phenomenon in a previously inactive galaxy. In this case, a compact disk does form out of the material from the disrupted star, and the material slowly accretes as well as spreads out, carrying the excess angular momentum. In this case, the disk is never actually stationary, even in the outer part, and in this case, our approach gives only a crude approximation. A new code, with different initial conditions and the outer (free) boundary condition, would be necessary for modeling, but at present, this is beyond the scope of our model. In this case, there is also no outer material, able to provide the BLR emission since to have these emission lines we need the central irradiation but also the copious gas there, at hundreds or thousands of $R_{Schw}$, ready to be ionized. 

The second scenario is the binary black hole (BBH) system, frequently invoked in CL AGN context anyway, and we present the schematic model of this scenario in Figure \ref{fig:tde_bbh}. If the smaller mass secondary black hole is already aligned with the accretion disk, it opens a gap in the disk, and the material still flows through the gap assisted by the secondary black hole and a small disk around it. In this case, the outer disk will provide the source of material for BLR, the inner edge of the gap will serve as $R_{out}$, and if the gap is not too broad it will not show very clearly in the broad band SED spectrum. in this case, the is a localized stream of material hitting the disk at $R_{out}$, like in galactic low mass X-ray binary systems, but apart from that the expected constant mass supply and constant value of  $R_{out}$ are well approximated in our numerical approach. 

We must stress that all presented results are based on a code that is based on the vertically-averaged structure of the disk as well as the corona, i.e. we have only two zones in the vertical direction. The frequent presence of the soft X-ray excess in AGN indicates the presence of the additional zone - the warm corona - which is an optically thick dissipating zone cooled predominantly by Comptonization \citep[e.g.][and the references therein]{czerny2003,rozanska2015,petrucci2020}. Recently, the full vertical structure of the disk including the effect of the magnetic field was calculated by \citet{gronkiewicz2020} for Galactic binaries, and it showed that the warm corona develops naturally in this case, but the disk is stabilized only for a certain range of radii. Similar behaviour is seen in AGN disks (Dominik Gronkiewicz, private communication). The code used by \citet{gronkiewicz2020}, however, is a stationary code so the predictions for the time-dependent behaviour in full 2-D are still to be developed. The available 3-D MHD global solutions do not cover the wide parameter range so they are not yet conclusive. Therefore, better modelling, as well as observational data of multiple outbursts across the whole black hole mass range, are needed to firmly establish the role of the radiation pressure instability in accreting black holes.

Our simulations show that the radiation pressure instability can explain some aspects of variability observed in accreting black holes across the broad mass range. It is interesting that a single mechanism can represent recurrent luminosity changes from microquasars to AGN, although in this last case a connection with TDE or a BBH has to be invoked to reproduce observed events in timescales of a few years or shorter.  However, as discussed in the Introduction, many different models of these phenomena were proposed, and they cannot be excluded. 

\section{Conclusions}
\label{sec:conclusions}

In this work, we explore the properties of outbursts for objects with 10, 10$^5$, and 10$^7 M_{\odot}$ based on radiation pressure instability using GLADIS code. We include accreting corona, inner ADAF, magnetic field presence, and the role of TDE in the system
 The main aspects of the modelling are the following:
 \begin{itemize}
     \item We show that we can obtain outbursts for 10, 10$^5$,  and 10$^7 M_{\odot}$ using the radiation pressure instability model.
     \item We confirm that the radiation pressure instability scenario can model heartbeat states in microquasars.
     \item To model proper timescales for Quasi-Periodic Ejection, we need an event at a relatively close distance to the central source and it may be explained by the TDE phenomenon.
    \item Repetitive outbursts in Changing Look AGN can be modelled with the use of radiation pressure instability including the cooling effect of magnetic field, however, this scenario also requires a small outer radius. It may be explained as TDE or with the presence of a gap in the disk due to the BBH.
     \item We notice that black hole mass is the leading parameter in dependence between period and magnetic field parameter.
     \item The assumption of a small outer radius of the accretion disk, for higher black hole masses (10$^5$ and 10$^7 M_{\odot}$) shortens the outbursts, but leads to complex multi-scale-amplitudes outbursts and resembles deterministic chaos behaviour.
          \item Strong magnetic field stabilizes the disk 
 \end{itemize}


\begin{acknowledgements}
The project was partially supported by the Polish Funding
Agency National Science Centre, project 2017/26/A/ST9/
00756 (MAESTRO 9), project 2021/41/N/ST9/02280 (PRELUDIUM 20), and MNiSW grant DIR/WK/2018/12.
BC wishes to thank ISSI for their kind hospitality for hosting the meeting over which some aspects related to this work were discussed, and in particular ISSI Warm Corona Team for helpful advice.
AJ was supported by grant 2019/35/B/ST9/04000 from Polish National Science Center.
\end{acknowledgements}

\bibliographystyle{aa}
\bibliography{aanda}

\end{document}